\documentclass[twocolumn]{article}

\pdfoutput=1

\usepackage[utf8]{inputenc}  
\usepackage{float} 
\usepackage{enumerate}  
\usepackage{bm}
\usepackage{amsmath} 
\usepackage{amsfonts}
\usepackage{mathrsfs} 
\usepackage{mathtools}
\usepackage{xcolor}
\usepackage{url}
\usepackage[left=0.75in,right=0.75in,top=1in,bottom=1in,%
            footskip=.25in]{geometry}

\usepackage[superscript,biblabel]{cite}

\definecolor{mygreen}{rgb}{0,0.834,0}

\begin{document}

\twocolumn[
  \begin{@twocolumnfalse}




    \begin{center}
      {\Large \textbf{Lagrangian diffusive reactor for detailed thermochemical computations of plasma flows}}
    \end{center}

    {\normalsize Stefano Boccelli$^{1,2,\,a)}$, Federico Bariselli$^{1,2,3,\,b)}$, Bruno Dias$^{1,4,\,c)}$
    and Thierry E.\ Magin$^{1,\,d)}$}
    
    \vskip2ex

    {\footnotesize
    $^{1}$ von Karman Institute for Fluid Dynamics, Aeronautics and Aerospace Department\\
    $^{2}$ Politecnico di Milano, Dipartimento di Scienze e Tecnologie Aerospaziali\\
    $^{3}$ Vrije Universiteit Brussel, Research Group Electrochemical and Surface Engineering\\
    $^{4}$ Universit\'e Catholique de Louvain, Department of Thermodynamics and Fluid Mechanics}

    \vspace{2ex}

    {\footnotesize
    $^{a)}$ stefano.boccelli@polimi.it\\
    $^{b)}$ federico.bariselli@vki.ac.be\\
    $^{c)}$ bruno.ricardo.barros.dias@vki.ac.be\\
    $^{d)}$ thierry.magin@vki.ac.be}
    

      \begin{abstract}
      The simulation of thermochemical nonequilibrium for the atomic and molecular energy level populations in plasma flows requires a comprehensive modeling of all the elementary collisional and radiative processes involved. Coupling detailed chemical mechanisms to flow solvers is computationally expensive and often limits their application to 1D simulations. We develop an efficient  Lagrangian diffusive reactor moving along the streamlines of a baseline flow simulation to {compute} 
      detailed thermochemical effects. In addition to its efficiency, the method allows us to model both continuum and rarefied flows, while including mass and energy diffusion.
      The Lagrangian solver is assessed for several testcases including strong {normal} 
      shockwaves, as well as 2D and axisymmetric blunt-body hypersonic rarefied flows.
      In all the testcases performed, the Lagrangian reactor improves drastically the baseline simulations. The computational cost of a Lagrangian recomputation is typically  orders of magnitude smaller with respect to a full solution of the problem.
      The solver has the additional benefit of being immune from statistical noise, which strongly affects the accuracy of DSMC simulations, especially considering minor species in the mixture. The results demonstrate that the method enables {applying} 
      detailed mechanisms to multidimensional solvers to study thermo-chemical nonequilibrium flows. 
      \vspace{1cm}
    \end{abstract}

  \end{@twocolumnfalse}
]





\vspace{1cm}

\section{Introduction}

A broad range of high-enthalpy and plasma technology applications exhibit thermochemical nonequilibrium effects, for instance in the fields of thermal plasmas,\cite{raiche_shock_nodate}  combustion, \cite{smith_gri-mech_nodate} plasma-assisted ignition,\cite{popov_kinetics_2016,anne2017} diagnostics,\cite{cooper_plasma_1966,lauxzare} solar physics,\cite{Maneva2017,Munafo2017} 
laser ablation, \cite{bulgakov_gas-dynamic_1998} surface coating, 
\cite{Shul1997} and in general, materials technology.\cite{Sanders1990} In aerospace applications,   the radiative heat flux to the heat shield of planetary entry probes \cite{helber_material_2016,taylor_monte_1994} depends on the populations of atomic and molecular internal energy levels, often out of equilibrium. In particular, many chemically reacting  species are present in atmospheric entry flows for space vehicles reaching  entry velocities higher than 10 km/s.\cite{park_review_1993, park_review_1994} With the ambition of deep space exploration,  detailed chemical mechanisms become of primary importance to optimize the efficiency of electric propulsion thrusters.\cite{croes_2d_2017,Sommerville2008}

The simulation of thermochemical nonequilibrium for the atomic and molecular energy level populations in plasma flows requires a comprehensive modeling of all the elementary collisional and radiative processes involved. Detailed simulations are based on a large set of chemical species and their related chemical mechanism.\cite{bultel_collisional-radiative_2006,munafo_modeling_2013,chaban_dissociation_2008,Bultel2013,pietanza_non-equilibrium_2018,le_complexity_2013} Coupling such mechanisms to flow solvers is computationally expensive and often limits their application to 1D simulations. Moreover, since the timescales of the physico-chemical phenomena vary widely,  the discretization of the convective and diffusive fluxes, as well as chemical production rates,                                                                                                                                                                                                                                                                                                             necessitates a specific numerical  treatment.\cite{young_numerical_1977,schwer_adaptive_2003,duartemassot}

The simulation of non-trivial geometries  becomes feasible by adopting                                                                               strategies to lower the computational cost associated with chemistry modeling, while retaining a good level of physical realm; principal component analysis (PCA),\cite{bellemans_reduction_2015} energy levels binning,\cite{le_complexity_2013, munafo_boltzmann_2014} and rate-controlled constrained-equilibrium (RCCE)\cite{keck_rate-controlled_1990} being some possible approaches.
Yet, the problem remains complicated enough to run out of the current supercomputing capabilities when real-world applications or design loops are targeted.
A pragmatic way to address the problem still consists in making strong simplifying assumptions.
%

In the chemistry and combustion communities, the idea of decoupling flow and chemistry is used to include detailed thermochemical effects into lower-fidelity baseline solutions,~\cite{corbetta_catalytic_2014} and also  to formulate hybrid Eulerian-Lagrangian reactors.\cite{raman_hybrid_2005}
In atmospheric entry plasmas, a Lagrangian method using a collisional-radiative reactor has been coupled to a flow solver,~\cite{magin_nonequilibrium_2006} based on the 1D method proposed by Thivet to study the relaxation of chemistry past a shockwave.\cite{thivet_modeling_1992} Lagrangian tools, allowing for the refinement of  an existing solution with very small computational effort, are  based on fluid models.

Fluid models for plasma flows can be derived from kinetic theory as asymptotic solutions to the Boltzmann equation, provided that the Knudsen number is small enough.\cite{cercignani_rarefied_2000,klimontovich_kinetic_1982,giovangigli_multicomponent_1999} For instance, Graille~$et~al.$~\cite{graille09} have obtained a perturbative solution for multicomponent plasmas based on a multiscale Chapmann-Enskog method by accounting for the disparity of mass between the electrons and heavy particles, as well as for the influence of the electromagnetic field. When the Knudsen number is too large to be in the continuum regime, the Maxwell transfer equations can be derived  by  directly averaging microscopic properties in the velocity space and by using the collisional operator properties.\cite{ferziger_mathematical_1972} This set of balance equations for mass, momentum, and total energy holds in both the continuum and rarefied regimes. The difficulty is then to find a closure for the transport fluxes appearing in the Maxwell transfer equations. A direct numerical simulation of the kinetic equations can be used to compute these fluxes.\cite{bird_molecular_nodate}

The aim of this work is to develop a method for including detailed thermochemical effects into lower-fidelity baseline solutions through an efficient  Lagrangian diffusive reactor. Our approach starts from a baseline solution for a plasma flow, by extracting the velocity and total density fields along its streamlines, and thus decoupling thermochemical effects from the flowfield. 
The main assumption  is that fine details of the species energy level populations, as well as trace species in the mixture, do not severely impact the hydrodynamic features of the flow. This is the case for many applications, such as the aerodynamics of a jet, the location of a detached shock wave, or the trail behind a body flying at hypersonic speed. As long as the total energy transfer can be modeled by means of some effective chemical mechanism, a more detailed description of the thermo-chemical state of the plasma can be obtained by re-processing the baseline calculation using more species and chemical reactions.
 The originality of this contribution consists in leveraging the Lagrangian nature of the proposed method, developing a general solution procedure based on an upwinded marching approach, adding rarefied, multi-dimensional, and dissipative effects.  The consequence is a drastic boost in the computational efficiency, allowing to deal with 
a large number of chemical species.

We propose to develop a tool  to obtain reasonably accurate predictions of the thermochemical state of a flow using an enlarged set of species and describing thermal nonequilibrium via multi-temperature, state-to-state, or collisional-radiative models. As a proof of concept, we account for radiation-flow coupling via escape factors. Reaching higher accuracies using detailed chemistry output of the Lagrangian reactor to solve the radiative transfer equation is beyond the scope of this work. 
The presented strategy can be employed as a design tool to obtain information too expensive for a fully coupled approach. Alternatively, it can also be used for diagnostic purposes to promptly estimate the effect of different physico-chemical models into realistic simulations: this allows us to understand whether a simplified modeling can be suitable for the considered problem. 
Finally, with respect to particle-based flow simulations, such as those obtained with the Direct Simulation Monte Carlo Method (DSMC),\cite{bird_molecular_nodate}  the proposed method smooth out the noise and irregularities associated to the inherent stochastic approach, improving the prediction of minor species. Electronic energy levels, which would require a particularly detailed and computationally intensive approach otherwise,\cite{li_modeling_2011} can also be easily computed.
The capabilities of the method are assessed against five problems, namely:
(i) Chemical refinement, (ii) Thermal refinement, 
(iii) State-to-state refinement, (iv) 2D rarefied  flow, and
(v) Mass and energy diffusion.
In all the performed testcases, we investigate the accuracy of the thermochemical description  and its computational cost.

Section \ref{sec:lagrangian-approach} of this work introduces the Lagrangian approach in terms of 
governing equations, physical modeling of fluxes and numerical implementation.
The method is applied to the five  testcases in section \ref{sec:results}.
Finally, in section \ref{sec:conclusions} some conclusions and future perspectives on the tool developed are drawn.  
\section{Lagrangian reactor approach}\label{sec:lagrangian-approach}

\begin{figure}
  \centering
  \includegraphics[width=0.8\columnwidth]{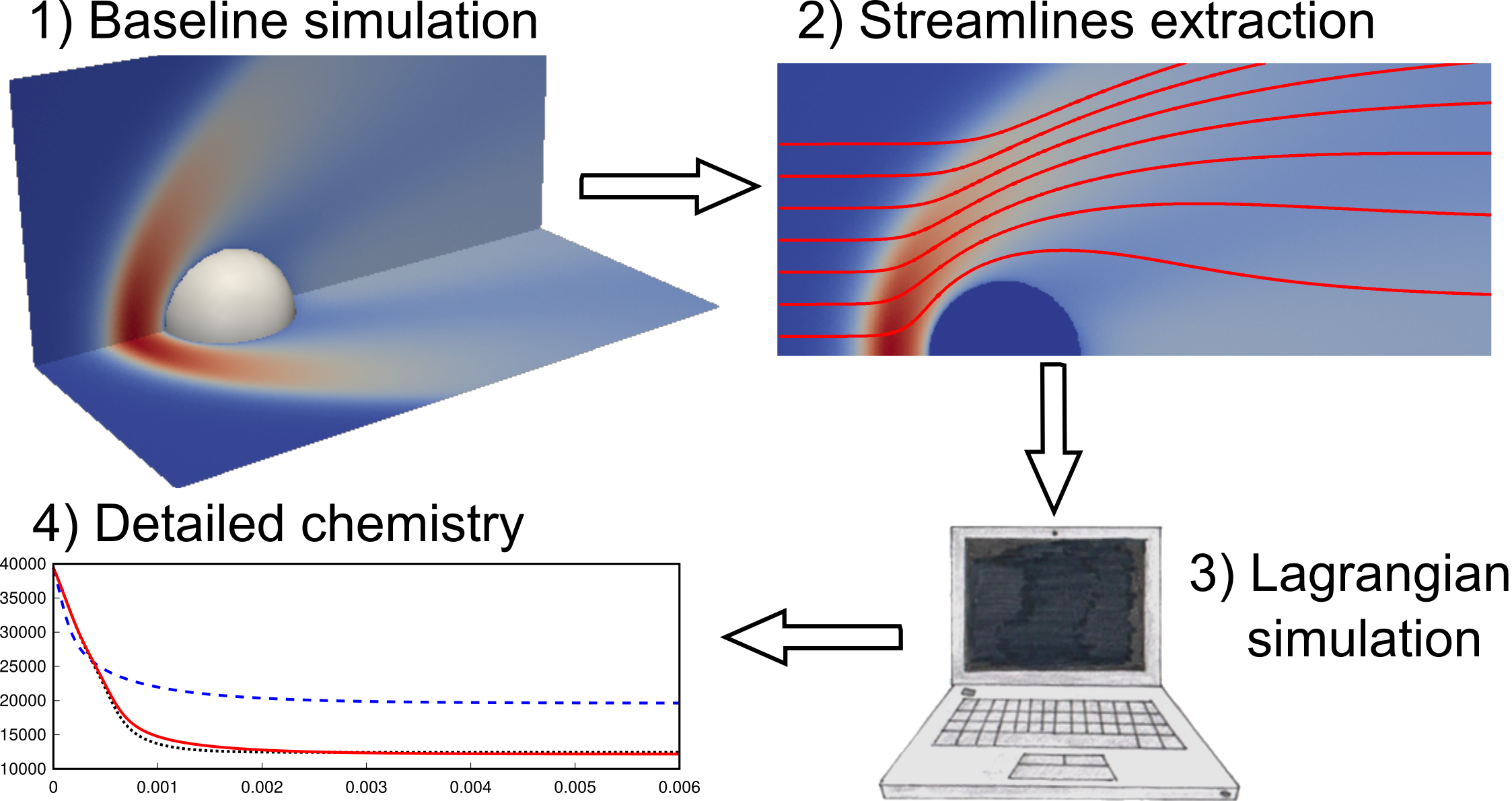}
  \caption{Procedure for applying the Lagrangian diffusive reactor to perform detailed thermochemical flow computations.}
  \label{fig:LARSEN-scheme}
\end{figure}

In this work, the flow is assumed to be steady, such that the trajectory of a fluid element coincides with its streamline. We propose to use this property in the development of a diffusive Lagrangian reactor moving along the streamlines.  The solution procedure, sketched in Fig.\ \ref{fig:LARSEN-scheme}, is as follows:
(i) A baseline flow simulation is obtained using an effective thermo-chemical model;
(ii) A suitable number of streamlines is extracted from the solution;
(iii) The Lagrangian tool imports the baseline streamlines and recomputes the species mass 
and energy transfer using more chemical species and thermo-chemical processes. This  variable volume chemical reactor follows 
a precomputed fluid element, and in the most elaborated approach developed, energy and mass are exchanged by diffusion with its neighbors over the computational domain.
The velocity field is not recomputed but taken from the baseline simulation. 
The global mixture density comes directly from imposing a velocity field and is thus
directly imported from the baseline simulation as well.
Initial conditions for the species concentration and temperatures are picked at the 
beginning of the streamlines, and the new thermo-chemical state is calculated by integration of species mass conservation equations and suitable energy equations given in this section.
It should be noted that there is no need for solving the momentum and global mass equations, since the velocity and density are imported.
These fields introduce information about the flow topology, as well as its kinetic energy. 



\subsection{Maxwell transfer equations and nonequilibrium models}\label{sec:MTE-equations-models}
%
For quasi-neutral plasmas and in the absence of applied electro-magnetic field, the Maxwell transfer equations are expressed as follows\cite{vincenti_introduction_1965, maxwell_iv._1867}
\begin{equation}
  \frac{\partial}{\partial t}
  \begin{pmatrix}
    \rho_i \\
    \rho \bm{u} \\
    \rho E 
  \end{pmatrix}
  +
  \bm{\nabla}\!
  \cdot \!
  \begin{pmatrix}
    \rho_i \bm{u} \\
    \rho \bm{u}\bm{u} \\
    \rho  \bm{u} H
  \end{pmatrix}
 + \bm{\nabla}\!
   \cdot \!
   \begin{pmatrix}
    \rho_i \bm{V}_i  \\
     P \mathbb{I}+\bm{\tau} \\
   \bm{\tau}\! \cdot\! \bm{u} + \bm{q}
  \end{pmatrix}
  =
  \begin{pmatrix}
  \omega_i \\
0
   \\
 \mathscr{P}
  \end{pmatrix},\label{eqMaxT}
\end{equation}
where index $i \in \mathcal{S}$ spans the chemical species.   The species partial density is introduced as $\rho_i$, together with the mixture density   
$\rho = \sum_{i\in \mathcal{S}} \rho_i$,   hydrodynamic velocity $\bm{u}$, total energy
$E$,  and total enthalpy $H$.  The transport fluxes are  obtained as average fluxes of microscopic properties in the velocity space weighted by the species velocity distribution function $f_i$, which is solution to the Boltzmann equation that can be computed, for instance, by means of a stochastic particle method.\cite{bird_molecular_nodate} The species diffusion velocities are then expressed as an average mass flux, $e.g.$, $\bm{V}_i=\int_ {\mathbb{R}^3} f_i m_i (\bm{c}_i-\bm{u})~\mathrm{d}\bm{c}_i$, for point particles $i\in \mathcal{S}$ without internal energy. 
In this contribution, the pressure tensor is assumed to be isotropic, quantity $P$ being the scalar mixture pressure and $\mathbb{I}$ the identity matrix; the reactive pressure is assumed to be negligible. 
Suitable expressions for the mixture viscous stress tensor  $\bm{\tau}$ and heat flux $\bm{q}$ can be found in references.\cite{chapmancowling,ferziger_mathematical_1972}  The $i$-species  chemical production rates $\omega_i$ are given by the law of mass action recalled in appendix. \cite{giovangigli_multicomponent_1999,parkbook} Quantity $\mathscr{P}$ is the radiative power of the plasma considered as a participating medium. When the Knudsen number is small enough to use the Chapman-Enskog method, the transport fluxes are shown to be linearly proportional to transport forces.  For instance, neglecting thermal diffusion, the diffusion velocities are modeled  by means of the generalized Fick law, $\bm{V}_i=-\sum_{j \in \mathcal{S}} D_{ij} ~\bm{d}_j$, $i\in \mathcal{S}$. The diffusion force is $\bm{d}_i=(\bm{\nabla} P_i -n_iq_i \bm{E})/P$,   with the partial pressure $P_i$, number density $n_i$, and charge $q_i$ of species $ i\in \mathcal{S}$, and the electric field $\bm{E}$.\cite{orlach} In this case,  the calculation of the species velocity distribution function is no longer required, since the multicomponent diffusion coefficients $D_{ij}$, $i,j\in \mathcal{S}$, have a closed form in terms of average cross-sections  based on binary interaction potentials between the species pairs. \cite{chapmancowling}

Electrons and heavy particles can exhibit distinct temperatures due to their mass disparity. A balance equation for the free electrons energy can also be derived from the Boltzmann equation by using an average electron kinetic energy in the velocity space
\begin{equation}
  \frac{\partial}{\partial t}
(    \rho_e e_e^t)
  +
  \bm{\nabla}\!
  \cdot \!(
    \rho_e e_e^t \bm{u})
+ P_e \bm{\nabla}\! \cdot \bm{u}
+\bm{\nabla}\! \cdot \bm{q}_e
=
\Omega_e
+ \mathscr{P}_e, \label{eqEE}
\end{equation}
where quantity $e_e^t$ is the translational energy of free electrons, $\bm{q}_e$, the electron heat flux, and $\mathscr{P}_e$, the electron radiative power.  The source term $\Omega_e$ describes the exchange of energy between the free electrons and heavy particles through elastic collisions, as well as through inelastic collisions driven by electrons. The latter correspond to chemical reactions ($e.g.$ electron-impact ionization) and elementary processes of excitation  for the internal energy modes of atomic and molecular species ($e.g.$ electron-impact excitation). The set of equations \eqref{eqMaxT}-\eqref{eqEE} allows for a state-to-state description of the plasma, considering the internal energy levels as pseudo-species. For instance, Cambier and Kapper \cite{kapper_ionizing_2011} have studied a mixture of species  comprising the electronic energy levels of a partially ionized argon gas for shock-tube experiments. 

Multi-temperature models assume that the populations of the internal energy levels of the atoms and molecules follow Maxwell-Boltzmann distributions. They require less basic data and computational resources than state-to-state models, but they can be inaccurate when nonequilibrium prevails.
In this framework, atomic and molecular internal
degrees of freedom (rotational, vibrational, and electronic energy levels) are grouped into several modes supposed to be thermal baths of energy $e^m$  at a distinct temperature.\cite{parkbook}
For each
 mode $m \in \mathcal{M}$ considered, an energy equation is expressed as  
\begin{equation}
  \frac{\partial}{\partial t}
  \left( \rho e^m \right)
+ \bm{\nabla}\!\cdot
  \left( \rho e^m \bm{u} \right)
  + \bm{\nabla}\! \cdot \bm{q}^m
= \Omega^m
+ \mathscr{P}^m,\label{eqEM}
\end{equation}
where $\mathcal{M}$ is the set of energy modes. The source term $\Omega^m$ describes the exchange of energy through chemical reactions and excitation of internal energy modes. Quantity $\mathscr{P}^m$ is the radiative power for mode $m$. Notice that eq.~\eqref{eqEE} has a structure similar to the one of eq.~\eqref{eqEM}, except for the term corresponding to the work of the electron compression force $P_e \bm{\nabla}\! \cdot \bm{u}$. The electron translational energy will thus be treated formally as an internal mode, separately or part of a thermal bath, depending on the ansatz followed (the electrons pressure term will be added accordingly). 

Finally, when all the energy levels are found to be in equilibrium with a single thermal bath, the problem is degenerate and  fully described by the  system~ \eqref{eqMaxT} for conservation of mass, momentum, and total energy.  For instance, Bruno and Giovangigli \cite{bruno} have studied the convergence of a two-temperature model towards a one-temperature model based on kinetic theory, focusing on the relaxation of internal temperature and the concept of volume viscosity.


\subsection{Governing equations for the Lagrangian reactor}\label{sec:lagrangian-governing-equations}

The hydrodynamic velocity $\bm{u}$ is assumed to be known from a baseline simulation. The mixture density $\rho$ comes directly from the mass conservation and is also carried along without being recomputed.
We recall that the diffusive Lagrangian reactor uses a more complete mechanism than in the baseline simulation. The set of chemical species can differ, as well as the model chosen to describe the population of the translational and internal energy levels, so that the species mass conservation equations need to be solved again. The translational temperatures of electrons and heavy particles and the temperatures of the internal energy baths, in the case of multi-temperature models, are computed using the relevant energy equations. The equations of Section \ref{sec:MTE-equations-models} are recast in Lagrangian  form along the streamlines.  Assuming steady state conditions, the directional derivative is transformed into the operator
$\bm{u} \cdot \bm{\nabla} \equiv U \tfrac{\partial}{\partial s}$, where $s$ is the 
curvilinear abscissa  along the streamline, and $U$, the module of the velocity vector.  
This system is expressed in the general form
\begin{equation}\label{eq:diffusive-Lagr-sys1}
\frac{\partial Q}{\partial s} 
+ A \ \bm{\nabla}\!\cdot (B\ \bm{\nabla} Q ) = f,
\end{equation}
where the source term vector is $f=f(Q)$. The transport fluxes are written in terms of gradients of the variables $Q$, based on a transport coefficient matrix $B(Q)$. The matrix $A(Q)$ comes from the transformation from conservative variables to the vector $Q$.

In the first equation of system \eqref{eqMaxT}, the species densities $\rho_i$ are substituted with the mass fractions
$Y_i = \rho_i / \rho$. Invoking global mass conservation, $i.e.$, $\bm{\nabla}\cdot  (\rho \bm{u})=0$, the species continuity equations 
are  obtained in Lagrangian form as follows\cite{anderson_hypersonic_2000}  
\begin{equation}\label{eq:mass-fractions-equation}
  \frac{\partial Y_i}{\partial s}  + \frac{1}{\rho U}\bm{\nabla}\! \cdot (\rho_i \bm{V}_i)
={\mathscr{S}_i^{Y} }
  \ , \quad i \in \mathcal{S},
\end{equation}
with  mass source terms $\mathscr{S}_i^{Y} 
  = 
\omega_i /(\rho U)$. The remaining equations depend on the choice of thermo-chemical model selected. Using the third equation of  system \eqref{eqMaxT}, a conservation equation for the total enthalpy is found
 \begin{equation}\label{eqHLag}  \frac{\partial H}{\partial s}  +\frac{1}{\rho U}\bm{\nabla}\! \cdot (\bm{\tau}\! \cdot\! \bm{u} +\bm{q})
=\mathscr{S}^{H},\end{equation}
with the total enthalpy source term
$\mathscr{S}^{H} 
=\mathscr{P}/(\rho U)$. 


In a {\bf state-to-state} approach, 
eq.~\eqref{eq:mass-fractions-equation} is written for the electrons and each energy level
of the  atoms and molecules, 
considered as separate  species $i\in \mathcal{H}$ in the mixture, where symbol $\mathcal{H} = \mathcal{S} \setminus \{e\}$ stands for this extended set of heavy particles.
The total enthalpy (per unit mass)  is thus split into five terms:
\smash{$H=\tfrac{1}{2}U^2+\sum_{j \in \mathcal{H}} (Y_j e_j^f+Y_j h_j^t)+Y_e e^f_e+Y_e h^t_e,~i.e.$}, (i) the flow kinetic energy;
heavy particles contribute through (ii) their internal energy, where quantity $\smash{e_i^f}$  is the formation contribution to the enthalpy of energy level
$i\in \mathcal{H}$,
(iii) their translational enthalpy, where $h_i^t$ is the translational contribution assumed to be at temperature $T$;
free electrons contribute through (iv) their  formation energy $e^f_e$,  and (v) their translational enthalpy $h^t_e$ evaluated at temperature $T_e$.
 A suitable Lagrangian form is derived from global mass conservation for the expression $
 P_e \bm{\nabla} \cdot \bm{u} =
  - U P_e \tfrac{\partial }{\partial  s}\ln \rho$ in eq.~\eqref{eqEE}, 
which can thus be computed along a streamline of the baseline solution.
 Hence, an equation for the free electron temperature $T_e$ is easily derived from eq.~\eqref{eqEE} 
\begin{equation}\label{eq:Te-s2s-lagrangian}
\frac{\partial T_e}{\partial s}  +\frac{1}{\rho_e U c_{v,e}}\left[\bm{\nabla}\! \cdot \bm{q}_e-e_e^t\bm{\nabla}\! \cdot (\rho_e\bm{V}_e)\right]
=
  \mathscr{S}^{T_e},
\end{equation}
where the electron temperature source term is introduced as
$\mathscr{S}^{T_e}  = (\mathscr{P}_e+\Omega_e+U P_e \tfrac{\partial }{\partial  s}\ln \rho-e_e^t\omega_e)/(\rho_e U c_{v,e})$
and the  specific heat at constant volume  for electrons as $c_{v,e} =\mathrm{d e_e^t}/ {\mathrm{d} T_e} $.  Then, using eq.~\eqref{eqHLag},
one gets  after some algebra an equation for the heavy-particle  temperature 
\begin{multline}\label{eq:T-s2s-lagrangian}
\frac{\partial T}{\partial s}  +\frac{1}{\rho U c_{p,h}^t}\left[\bm{\nabla}\! \cdot (\bm{\tau}\! \cdot\! \bm{u}+\bm{q}-\gamma_e\bm{q}_e)\right.
\\-\sum_{j\in\mathcal{S}}(h_j^t+e_j^f)\bm{\nabla}\! \cdot (\rho_j\bm{V}_j)+\gamma_e e_e^t\bm{\nabla}\! \cdot (\rho_e\bm{V}_e)]
= \mathscr{S}^{T},
\end{multline}
where the heavy-particle temperature source term is \smash{$\mathscr{S}^{T}=
[\mathscr{P}-\gamma_e(\mathscr{P}_e+\Omega_e+U P_e \tfrac{\partial}{\partial  s}\ln{\rho})- \tfrac{1}{3}\rho \frac{\mathrm{d} }{\mathrm{d}s}U^3-$} \smash{$\sum_{j\in\mathcal{S}}(h_j^t+e_j^f)\omega_j+\gamma_ee_e^t\omega_e]/(\rho U c_{p,h}^t)$}, with the electron specific heat ratio $\gamma_e=c_{p,e}/c_{v,e}$ and specific heat at constant pressure   \smash{$c_{p,e} =\mathrm{d} h_e^t/ {\mathrm{d} T_e} $}. The translational contribution of the frozen mixture specific heat at constant pressure  \smash{$ c_{p,h}^t=\sum_{j\in\mathcal{H}} Y_j c_{p,j}^t$} is based on quantity \smash{$c_{p,i}^t={\mathrm{d} h_i^t } / {\mathrm{d} T}$} for species $i\in\mathcal{H}$. 
To summarize, in the state-to-state approach, one solves eqs.~\eqref{eq:mass-fractions-equation}, \eqref{eq:Te-s2s-lagrangian}, and \eqref{eq:T-s2s-lagrangian}. The vector of unknowns for system \eqref{eq:diffusive-Lagr-sys1} is $Q$~$=$~$[(Y_i)_{i\in\mathcal{S}}, ~T, ~T_e]^{\rm T}\in \mathbb{R}^n$ where the  dimension $n=N+2$, with the number of species $N={\rm card}(\mathcal{S})$.

In a {\bf multi-temperature} approach, the internal energy of mode $m$  is introduced as $\sum_{j \in \mathcal{S}} Y_j e_j^m$, where quantity $e_i^m$ is the mode  energy for species $i$ evaluated at temperature $T^m$.
An equation for the mode temperature  is derived from eq.~\eqref{eqEM}
\begin{equation}\label{eq:TML}
  \frac{\partial T^m}{\partial s}+\frac{1}{\rho  c_{v}^m U}\left[\bm{\nabla}\! \cdot \bm{q}^m-\sum_{j\in\mathcal{S}}e_j^m\bm{\nabla}\! \cdot (\rho_j\bm{V}_j)\right]
= \mathscr{S}^{T^m},
\end{equation}
where $m\in\mathcal{M}$. The source term for temperature is introduced as  {$\mathscr{S}^{T^m} =(\mathscr{P}^m+\Omega^m+\aleph_{e}^m U P_e \tfrac{\partial}{\partial  s}\ln{\rho}-\sum_{j\in\mathcal{S}}e_j^m\omega_j)/(\rho U c_{v}^m)$}. Quantity $c_{v}^m= \sum_{j \in \mathcal{S}} Y_j c_{v,i}^m,$ is the contribution of the frozen mixture specific heat at constant volume for mode $m$. 
    Quantities $e_i^m$ and  $c_{v,i}^m=\mathrm{d} e_i^m/ {\mathrm{d} T^m} $ are respectively  the energy and specific heat of mode $m$ for species $i \in\mathcal{S}$. 
When the translation of free electrons is also included in the mode, $i.e.$, $e^m_e=e^t_e$ with $T_e=T^m$, eqs. \eqref{eqEE} and \eqref{eqEM} need to be summed before recasting  them in the Lagrangian form given by eq.~\eqref{eq:TML}. The last equation is also valid when the thermal bath describes only the free electron translation, leading to the same structure as the one of  eq.~\eqref{eq:Te-s2s-lagrangian}. In both cases, quantity $-P_e \bm{\nabla}\! \cdot \bm{u}$ is added to the source term by means of symbol $\aleph_{e}^m=1$ if electrons translational energy belongs to pool $m$, otherwise zero.
Then, using eq.~\eqref{eqHLag}, one gets an equation for the translational mode of heavy particles 
\begin{multline}\label{eq:T-MT-lagrangian}
\frac{\partial T}{\partial s}  +\frac{1}{\rho U c_{p,h}^t}[\bm{\nabla}\! \cdot (\bm{\tau}\! \cdot\! \bm{u}+\bm{q})-\bm{\nabla}\! \cdot \sum_{m\in\mathcal{M}}\gamma^m \bm{q}^m
\\-\sum_{j\in\mathcal{S}}h_j\bm{\nabla}\! \cdot (\rho_j\bm{V}_j)+{\displaystyle \sum_{\scriptsize
\begin{array}{c} m\in\mathcal{M}\\ j\in\mathcal{S}\end{array}} }\gamma^m  e_j^m \bm{\nabla}\! \cdot (\rho_j\bm{V}_j)]
= \mathscr{S}^{T},
\end{multline}
where the heavy-particle temperature source term is
\smash{$\mathscr{S}^{T}=
[\mathscr{P}-\sum_{m\in\mathcal{M}}\gamma^m(\mathscr{P}^m+\Omega^m+\aleph_{e}^m U P_e \tfrac{\partial}{\partial  s}\ln{\rho})-$} \smash{$\tfrac{1}{3}\rho \frac{\mathrm{d} }{\mathrm{d}s}U^3-\sum_{j\in\mathcal{S}}h_j\omega_j+\sum_{j\in\mathcal{S},m\in\mathcal{M}}\gamma^m e_j^m\omega_j]/(\rho U c_{p,h}^t)$},
with the species enthalpies \smash{$h_i=h_i^t+\sum_{m\in\mathcal{M}}h_i^m+e_i^f$}, $i\in\mathcal{H}$, and \smash{$h_e=h_e^t+e_e^f$}, the  ratio of frozen specific heats   $\gamma^m={c_{p}^m}/{c_{v}^m}$  for mode $m$ and its contribution to the frozen mixture specific heat at constant pressure $c_{p}^m= \sum_{j \in \mathcal{S}} Y_j c_{p,i}^m$.  Quantity $c_{p,i}^m=\mathrm{d} h_i^m/ {\mathrm{d} T^m}$ is the specific heat at constant pressure of mode $m$ for species $i \in\mathcal{S}$. Notice the equality $c_{p,i}^m=c_{v,i}^m$, for heavy particles $i\in\mathcal{H}$.
When some of the atomic and molecular internal degrees of freedom  are grouped with the translational mode of heavy particles at the same bath temperature $T$, the expressions for the translational contribution of the enthalpy  $h_{i}^t$ and specific heat   $c_{p,i}^t$, $i\in\mathcal{H}$, used in Eq.~\eqref{eq:T-MT-lagrangian}, need to be modified accordingly. In the multi-temperature approach, one solves eq.~\eqref{eq:mass-fractions-equation},  one eq. \eqref{eq:TML}  
for every mode $m$ considered, and eq.~\eqref{eq:T-MT-lagrangian}. The vector of unknowns is $Q$~$=$~$[(Y_i)_{i\in\mathcal{S}}, ~T, ~(T^m)_{m\in\mathcal{M}}]^{\rm T} \in \mathbb{R}^n$, where dimension $n=N+M+1$, with the number of modes $M={\rm card}(\mathcal{M})$.

Finally, {\bf thermal equilibrium} is obtained as a degenerate case of the multi-temperature approach
\begin{multline}\label{eq:T-ST-lagrangian}
\frac{\partial T}{\partial s}  +\frac{1}{\rho U c_{p}}[\bm{\nabla}\! \cdot (\bm{\tau}\! \cdot\! \bm{u}+\bm{q})-\sum_{j\in\mathcal{S}}h_j\bm{\nabla}\! \cdot (\rho_j\bm{V}_j)]
= \mathscr{S}^{T},
\end{multline}
where the heavy-particle temperature source term is
\smash{$\mathscr{S}^{T}=
[\mathscr{P}-\tfrac{1}{3}\rho \frac{\mathrm{d} }{\mathrm{d}s}U^3-\sum_{j\in\mathcal{S}}h_j\omega_j]/(\rho U c_{p})$},
with the contribution to the frozen mixture specific heat at constant pressure $c_{p}= \sum_{j \in \mathcal{S}} Y_j c_{p,j}$. The species enthalpy $h_i$, and specific heat at constant pressure $c_{p,i}$, $i\in\mathcal{S}$, are functions of a single temperature $T$, gathering all the translational and internal modes of the species, as well as their formation enthalpy. In the thermal equilibrium approach, one solves eqs.~\eqref{eq:mass-fractions-equation} and \eqref{eq:T-ST-lagrangian}. The vector of unknowns is $Q$~$=$~$[(Y_i)_{i\in\mathcal{S}}, ~T]^{\rm T} \in \mathbb{R}^n$, where dimension $n=N+1$.

\subsection{Solution of the Lagrangian equations}\label{subsec:sol-Lagrangian-eq}
In system \eqref{eq:diffusive-Lagr-sys1}, the structure of the diffusion term is of elliptic nature, which in principle requires to solve all the points of the domain simultaneously.
In this work, two formulations are investigated to drastically reduce the computational cost: a \textit{decoupled single-streamline} mode and a 
\textit{coupled multi-streamline} mode.

\vspace{1ex}
\noindent \textbf{Decoupled single-streamline mode.} In the absence of diffusive terms in eq.~\eqref{eq:diffusive-Lagr-sys1}, the system of Lagrangian equations simplifies as follows, $ {\mathrm{d} Q}/{\mathrm{d} s} =f$, with the  vector of unknowns $Q=Q(s)\in \mathbb{R}^n$. It is a 
system of explicit ordinary differential equations of order one and dimension $n$. 
This structure is of great benefice from the computational point of view since it allows to retrieve the solution by marching along the streamlines.
In this case, the values of $Q$ can be computed along one independent streamline starting  from an initial value $Q|_{s=0}=Q_0$, while the streamline shape can develop in a 3D space.
Integration is performed  using either an adaptive Runge-Kutta 4-5 or implicit Rosenbrock 4 scheme, from the \texttt{C++} Boost library.\cite{noauthor_boost_nodate}

A major drawback of this approach is that the fluid element is assumed to be adiabatic.
This assumption may be acceptable considering the magnitude of the transport fluxes in advection-dominated (high Péclet number) flows.
However, across shock waves, near non-adiabatic walls and, more generally, in flows with strong shear and fluxes, this assumption breaks down; one needs a way to compute diffusion fluxes through the fluid element.

A possible remedy consists in importing the variations of total enthalpy directly from the baseline solution.
The diffusive fluxes are extracted from eq.~\eqref{eqHLag} as follows 
\begin{equation}
  \frac{1}{\rho U}\bm{\nabla}\! \cdot (\bm{\tau}\! \cdot\! \bm{u} +\bm{q})=
\frac{{\mathscr{P}}^*}{\rho U} -\frac{\Delta H^*}{\Delta s} 
\end{equation}
where symbol $^*$ refers to the baseline solution.
The fluxes employed with this approach are not self-consistent, since they are entirely
taken from the baseline flow, and not obtained from the recomputed solution.
This approach should thus be considered more as a useful engineering ploy, rather than 
a rigorous treatment.
Although not self-consistent, it has an interesting consequence:
extending the applicability range of the Lagrangian reactor towards conditions of significant 
rarefaction, despite its fluid formulation.
In fact, as long as the baseline solution is obtained with a  method  compliant with the rarefied regime
(such as the DSMC method), the Lagrangian reactor imports such fluxes without the need of 
imposing any preassigned model. 

\vspace{1ex}
\noindent \textbf{Coupled multi-streamline mode.} 
The second approach analyzed in this work aims at describing flows that show a preferential direction of development, such as boundary layers, jets, and trails, where diffusion mainly acts transversely. Calling $x$ the preferential  and $r$ the transverse direction (orthogonal to it), diffusion terms in eq.\ \eqref{eq:diffusive-Lagr-sys1} become purely transverse
\begin{equation}\label{eq:diffusive-Lagr-sys-transv}
  \frac{\partial Q}{\partial s} 
+ A \frac{\partial} {\partial r} \left(B \frac{\partial Q}{\partial r} \right) 
= f,
\end{equation}
with the vector $Q=Q(r,s)\in \mathbb{R}^n$. 
%
%
Additionally, the preferential direction $x$ is chosen as a common reference to all the streamlines. The material derivative is recast  from the curvilinear abscissa $s$ to the direction $x$ by introducing the local slope $\alpha_k$,  such that $\mathrm{d}x = \cos{\alpha_k}\, \mathrm{d}s_k$, $k\in \mathcal{K}$, where $\mathcal{K}$ is the set of indexes for the streamlines.
In this way, their phase is not lost during the integration (due to streamlines having different velocities) and governing equations can be integrated simultaneously, evaluating transverse diffusion terms implicitly at the current integration step, using a second-order central discretization.
The parabolic system \eqref{eq:diffusive-Lagr-sys-transv}, written for each streamline $k$, becomes 
\begin{equation}\label{eq:diffusive-Lagr-sys-transv-along-x}
  \frac{\partial Q}{\partial x} 
+ \frac{A}{\cos{\alpha_{k}}} \frac{\partial} {\partial r} \left(B \frac{\partial Q}{\partial r} \right) 
= \frac{f}{\cos{\alpha_{k}}},
\end{equation}


The $x$ integration is performed as for the decoupled single-streamline mode with adaptive Runge-Kutta 4-5 or implicit Rosenbrock 4 schemes.
Diffusion terms are discretized using a finite volume 1D formulation, where the cells are centered along the streamlines.
Diffusion fluxes and source terms are evaluated implicitly by the routine, at every integration step.
The simulation starts by imposing initial conditions $Q|_{s=0}=Q_0$ at the beginning of the streamline.
Transverse fluxes evaluation requires imposing boundary conditions  on the first and last streamlines, which can be either of Dirichlet (thermodynamic state imposed) or Neumann  (zero gradient) type.
In axisymmetric configuration, fluxes are imposed by symmetry on the lowest cell interface, located at the axis ($r = 0$).
\begin{figure}[htpb]
  \centering
  \includegraphics[width=0.6\columnwidth]{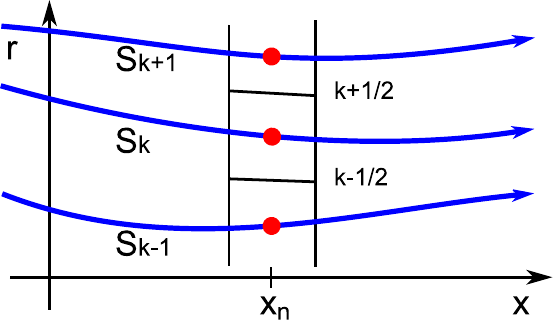}
  \caption{Streamline marching scheme with implicit transverse finite volume method.}
  \label{fig:scheme-FV-transverse}
\end{figure}
The physico-chemical properties are obtained through the Open Source 
\texttt{Mutation++} library.\cite{scoggins_development_2014, noauthor_mpp_nodate}

The implementation of the coupled multi-streamline Lagrangian approach in a 
\texttt{C++} language code has been verified by comparison of the analytical solution to a scalar diffusion equation \smash{$ {\partial Q}/{\partial s} =\alpha ~{\partial^2Q}/{\partial r^2}$}, where quantity $\alpha$ is a nonnegative constant. The problem of diffusion in a semi-infinite domain can be shown to have a self-similar solution, the similarity variable
being $r/\sqrt{4\alpha x}$, and its solution is given in terms of the Gauss error function.\cite{lienhard2013heat}
Given a maximum value for the independent variable $x_{max} $, parallel streamlines were created at positions from $r=0$ up to $r=10\sqrt{4\alpha x_{max}}$, where fluxes are negligible.
The thermodynamic state $Q$ was imposed on the bottom streamline and null fluxes (Neumann BC) on the top one. Using about 20 streamlines, with a progressively increasing spacing, allowed to retrieve the analytical solution within a few $\%$ accuracy. 
Increasing the number of streamlines reproduces the exact solution.

\section{Results}\label{sec:results}

In this section, the Lagrangian approach is assessed for five types of problems.
The chemical (A) and thermal (B) refinement capabilities are tested in the continuum regime for  the thermo-chemical relaxation  of an air mixture past a strong 1D shockwave, using both single and multi-temperature models. The state-to-state refinement (C) is analyzed in the continuum regime for the excitation and ionization of an argon plasma past a strong 1D shockwave, then applied to the rarefied regime based on a 2D simulation using the DSMC method and accounting  for radiative processes (D). Finally, mass and energy  diffusion capabilities are tested in air for the rarefied regime along the trail past a blunt body moving at hypersonic speed (E).


\subsection{Chemical refinement}
We study the chemical relaxation in air past a strong 1D shockwave by assuming thermal equilibrium.
Freestream conditions representative of the trajectory point at 1636~s of the Fire II flight experiment  \cite{cauchon1967radiative, panesi2009fire} are given in Table \ref{tab:fs-air5-air11-1T}.
\begin{table}[htpb]
\centering
\begin{tabular}{ccc}
$V_{\infty}$ [m/s] & $P_{\infty}$ [Pa] & $T_{\infty}$ [K] \\
\hline
11\,310 & 5.25 & 210 \\
\end{tabular}
\caption{Freestream conditions for Fire II trajectory point at 1636~s (Testcases A and B).}
\label{tab:fs-air5-air11-1T}
\end{table}

Two mixtures are considered in this section:  a neutral mixture composed of 5 species denoted by the set 
$\mathcal{S}_5 = \{\mathrm{N_2}$, $\mathrm{O_2}$, $\mathrm{NO}$, $\mathrm{N}$, $\mathrm{O}\}$, so-called ``air-5," and  a partially ionized mixture composed of 11 species denoted by the set $\mathcal{S}_{11} = \{ \mathrm{N_2}$, $\mathrm{O_2}$,  $\mathrm{NO}$, $\mathrm{N}$, $\mathrm{O}$, $\mathrm{N_2^+}$, $\mathrm{O_2^+}$,  $\mathrm{NO^+}$, $\mathrm{N^+}$, $\mathrm{O^+}$, $\mathrm{e^-}\}$, so-called ``air-11."
The chemical mechanism and rate coefficients are taken from Park~$et~al.$ \cite{park_chemical-kinetic_2001} The initial composition is assumed to be in local thermodynamic equilibrium at the freestream conditions.
First, the post-shock mass, momentum, and total energy are computed using the Rankine-Hugoniot relations, keeping the chemical composition frozen. The baseline solution is obtained by solving numerically the system \eqref{eqMaxT} for the air-5 mixture in steady conditions, following the approach proposed by Thivet.\cite{thivet_modeling_1992}
The solution in the post-shock region is then recomputed by means of the Lagrangian reactor, which imports the velocity and density fields. The Lagrangian  eqs.~\eqref{eq:testcase-A-mass} and \eqref{eq:testcase-A-temperature} presented in Appendix are  solved using the air-11 mixture.
Finally, the results of the Lagrangian simulation are compared to a reference solution to system \eqref{eqMaxT} obtained for the air-11 mixture.

\begin{figure}
  \centering
  \includegraphics[width=1.0\columnwidth]{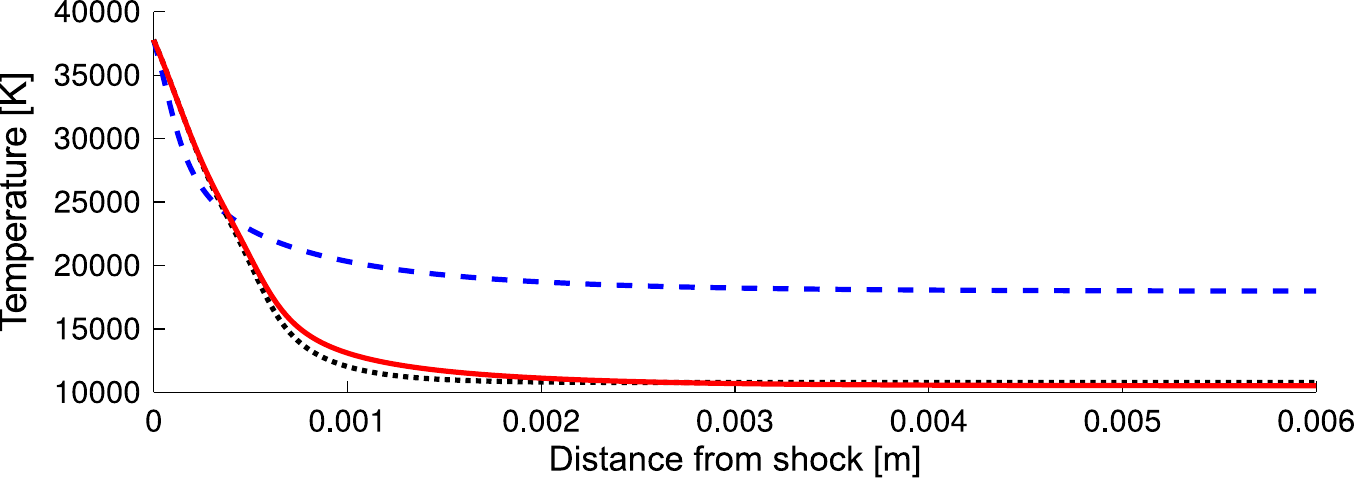}
  \caption{Chemical refinement for Fire II-Testcase A: temperature in the post-shock region; \textcolor{blue}{\bf -\,-\,-} baseline simulation (air-5); \textcolor{red}{\bf -----} Lagrangian simulation (air-5 $\rightarrow$ air-11); $\bm{\cdots}$ reference simulation (air-11).}
  \label{fig:air5-air11-1T-temp}
\end{figure}
\begin{figure}
  \centering
  \includegraphics[width=1.0\columnwidth]{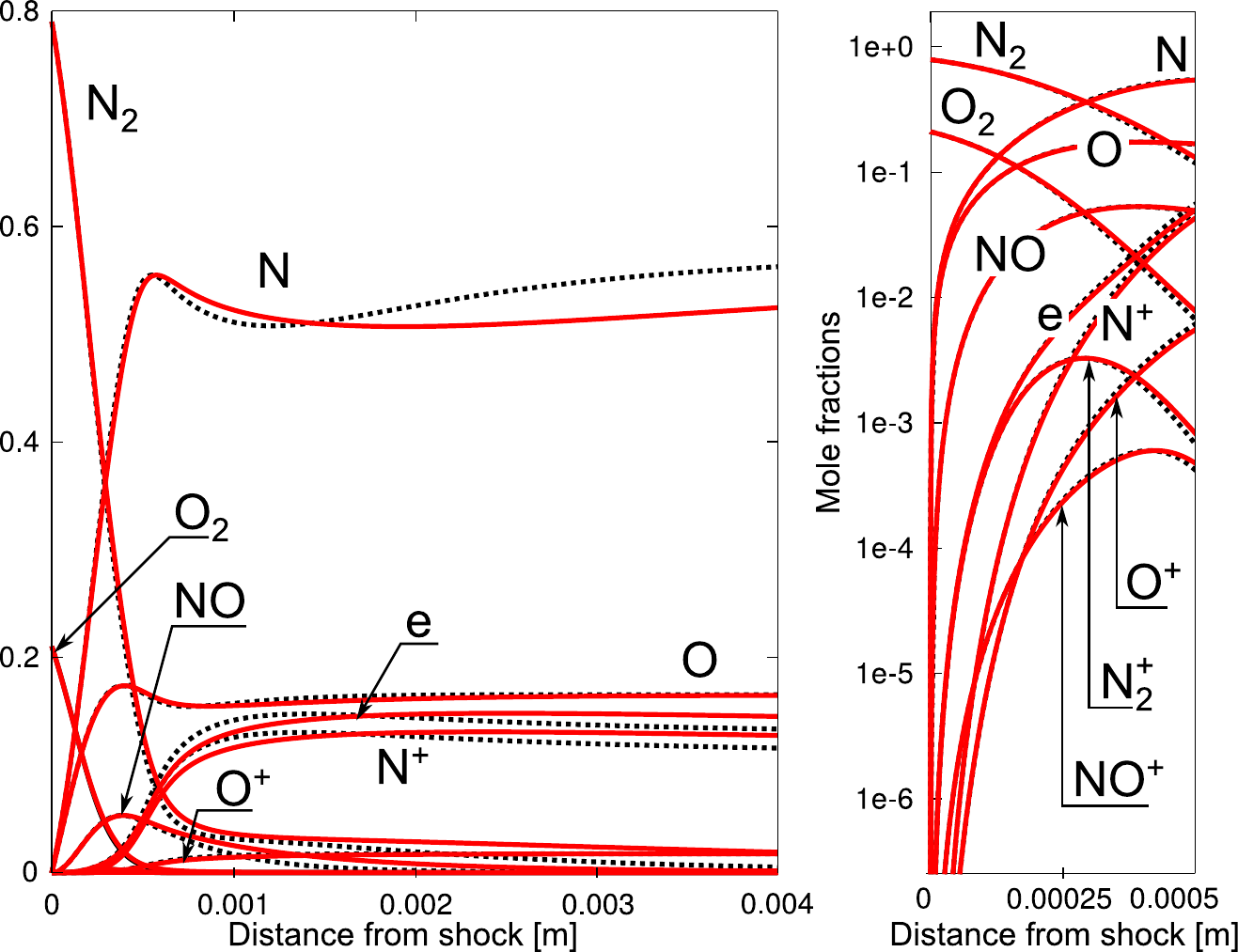}
  \caption{Chemical refinement for Fire II-Testcase A: species mole fractions in the post-shock region. Left: linear scale; right: logarithmic scale, detail of near-shock region;
  \textcolor{red}{\bf -----} Lagrangian simulation (air-5 $\rightarrow$ air-11); $\bm{\cdots}$ reference simulation (air-11).
  }
  \label{fig:air5-air11-1T-species-all}
\end{figure}

The temperature and 
species concentration along the stagnation line are shown in Figs.\ \ref{fig:air5-air11-1T-temp} and 
\ref{fig:air5-air11-1T-species-all}.
The Lagrangian simulation results eventually reach equilibrium values in agreement with the reference simulation. It is noteworthy that ions and free electrons, not present in the baseline solution, 
are reproduced with good accuracy.
The discrepancies between the Lagrangian and reference results start around the position 
$x = 5\times 10^{-4} \, \textrm{m}$, where the bulk of ionization occurs. In the reference computation, energy is spent to ionize the species; the kinetic energy 
is thus reduced.  The error originates from the derivative of the kinetic energy flux $U^3$, appearing in eq.~\eqref{eq:testcase-A-temperature}. However, the results are in fair agreement. Notice that the mass flux $\rho U$ is an invariant of all the simulations; this term, appearing at the denominator of the source term of the Lagrangian solution, does not introduce any approximation. 

This testcase shows the chemical refinement capabilities of the method: the temperature field departs from the baseline solution shortly after the shock and, together with the chemical composition, closely follows the reference result.
In general, the smaller the influence of the new chemical model on 
the baseline energy balance, the greater the expected accuracy.
If only trace species are introduced, small degrees of ionization or dissociation  almost do not alter the energy 
balance and the method retrieves the right solution with a very good degree of accuracy. However, this criterion is much less strict than it seems:
in the simulation presented, newly-introduced ions reach concentrations over 10\%, yet the solution 
is reproduced with reasonable accuracy.


\subsection{Thermal refinement}
We study the thermo-chemical relaxation in air past a strong 1D shockwave for the air-5 mixture, allowing for thermal nonequilibrium to occur. The translational energy of heavy particles and rotational energy of molecules are assumed to be in equilibrium at temperature $T$, while the vibrational energy of molecules and electronic energy of heavy particles are at temperature $T^v$. The freestream pressure and temperature are identical to the ones of Testcase A, while the velocity was reduced to $7000$ m/s to better comply with the air-5 model.
With the same spirit as in the previous section, the solution  is  recomputed in the post-shock region by means of the Lagrangian reactor using a two-temperature ($T$-$T^v$) model, starting from a single-temperature  baseline solution. 
 The Lagrangian  eqs.~\eqref{eq:testcase-A-mass}, \eqref{eqAp1}, and \eqref{eqAp2} presented in Appendix are  solved using the same air-5 mixture.
A reference solution is obtained for the air-5 mixture by solving in steady conditions the system \eqref{eqMaxT} coupled to eq.~\eqref{eqEM} for the mode at temperature $T^v$,  following the approach proposed by Magin~$et~al.$~\cite{magin_nonequilibrium_2006}

The single-temperature post-shock conditions used for the baseline solution satisfy thermal equilibrium, whereas the two-temperature conditions used for the reference solution assume that the vibrational-electronic mode is frozen, so that the translational temperature $T$ jumps, while the internal temperature $T^v$ remains 
at its freestream value.
The two temperatures then relax towards each other and eventually equilibrate at the same value 
reached by the single-temperature solution.

\begin{figure}
     \centering
     \includegraphics[width=1.0\columnwidth]{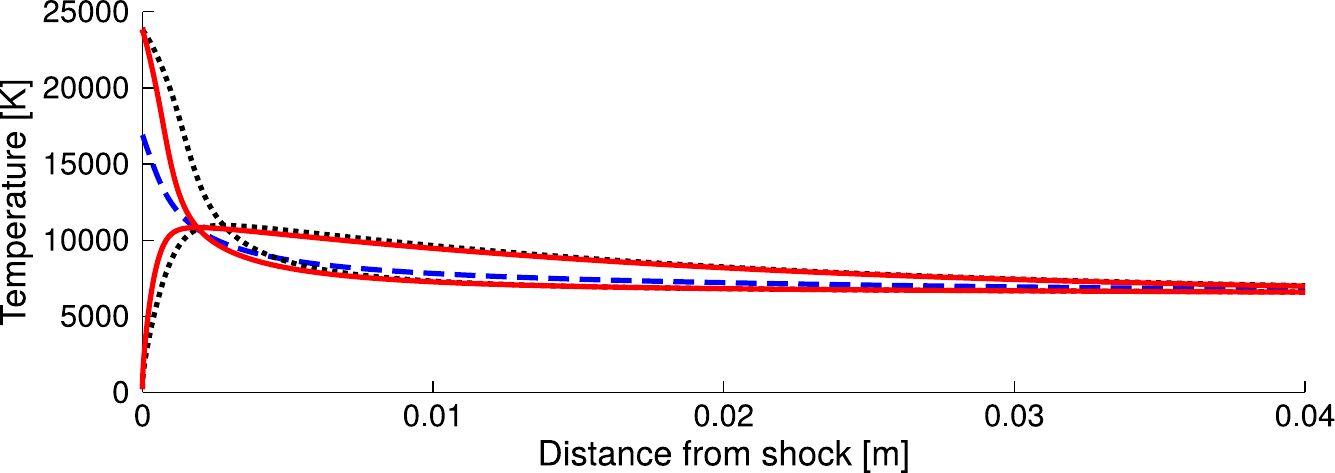}
     \caption{Thermal refinement for Fire II-Testcase B: translational and internal temperatures in the post-shock region; 
     \textcolor{blue}{\bf -\,-\,-} baseline simulation (1$T$); \textcolor{red}{\bf -----} Lagrangian simulation (1$T$ $\rightarrow$ $T$-$T^v$); $\bm{\cdots}$ reference simulation ($T$-$T^v$).}
     \label{fig:air5-1T-TTv-temp}
   \end{figure}

\begin{figure}
     \centering
     \includegraphics[width=1.0\columnwidth]{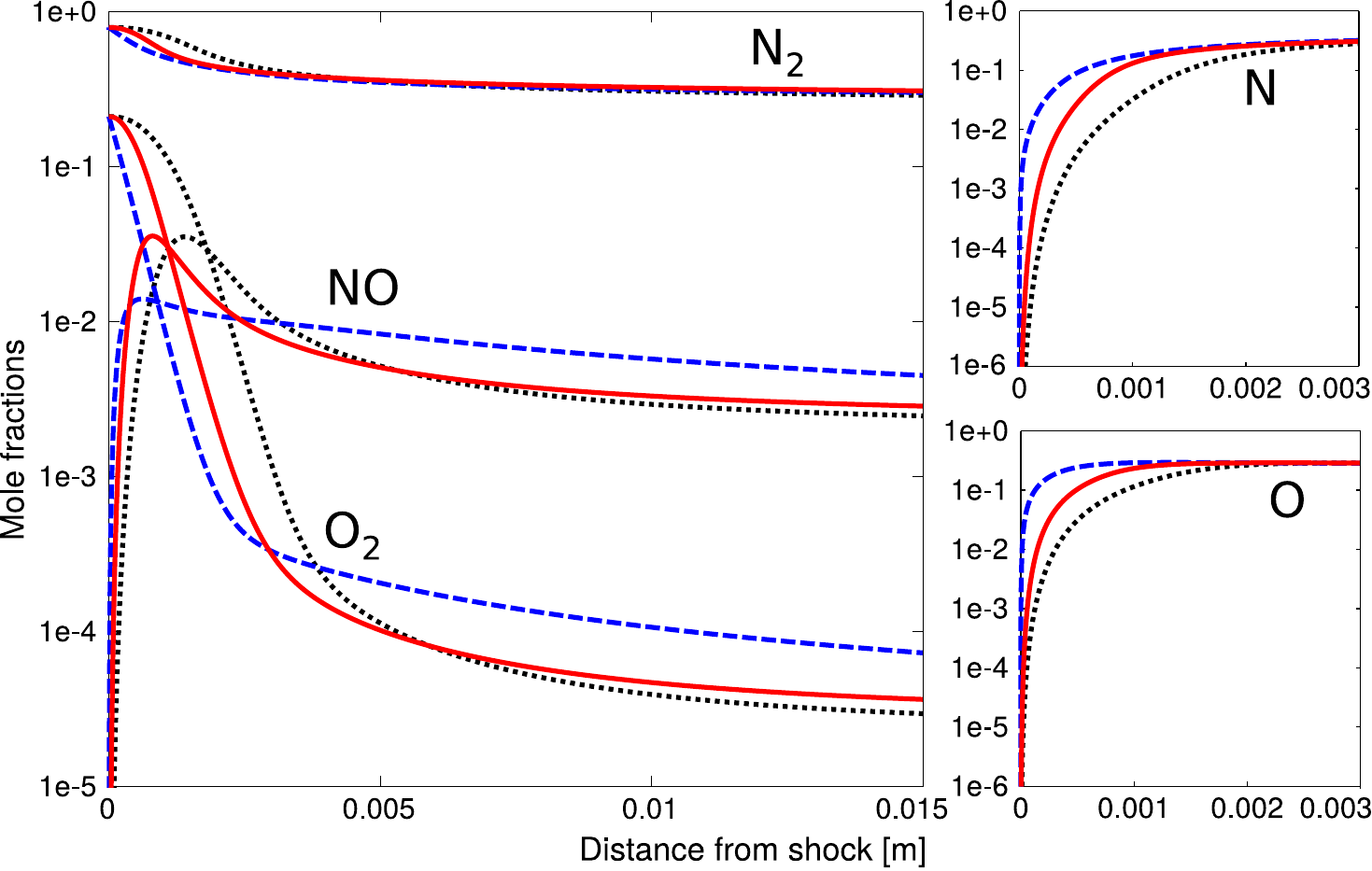}
     \caption{Thermal refinement for Fire II-Testcase B: species mole fractions in the post-shock region;
         \textcolor{blue}{\bf -\,-\,-} baseline simulation (1$T$); \textcolor{red}{\bf -----} Lagrangian simulation (1$T$ $\rightarrow$ $T$-$T^v$); $\bm{\cdots}$ reference simulation ($T$-$T^v$).} 
     \label{fig:air5-1T-TTv-spec}
   \end{figure}

The Lagrangian solution takes the velocity and density fields from the single-temperature baseline
computation, but starts with temperatures out of equilibrium.
The solution is shown in Figs.\ \ref{fig:air5-1T-TTv-temp} and 
\ref{fig:air5-1T-TTv-spec}.
The  temperatures obtained by means of the Lagrangian solver follow closely the reference solutions, except for a slight 
anticipation of the point where translational and internal temperatures cross each other.
Concerning the chemical composition, a couple of important points should be highlighted:
(1) The Lagrangian solution is improved by at least  40\% and 
approaches the reference result much quicker than the baseline solution; 
(2) The peak of $\mathrm{NO}$ is slightly anticipated in position but 
correctly reproduced in amplitude, while it was originally underpredicted by the baseline solution.
A correct prediction of this peak is crucial to an accurate calculation of the shock layer radiation.\cite{soucasse}


\subsection{State-to-state refinement}
We study the excitation and ionization 
of an argon plasma in the continuum regime past a strong 1D shockwave, using  a state-to-state mechanism. The freestream conditions for the shock-tube case studied, representative of an experiment performed
at the University of Toronto’s Institute of Aerospace Studies (UTIAS), \cite{glass} are given in Table \ref{tab:utias-case}.
\begin{table}[htpb]
\centering
\begin{tabular} {cccc}
$V_\infty$ [m/s] & $P_\infty$ [Pa] & $T_\infty$ [K] &  $\rm{Ma}_\infty$ [-] \\
\hline
5\,103 & 685.3 & 293.6 & 15.9 \\
\end{tabular}
\caption{Freestream conditions for the UTIAS argon shock-tube experiment (Testcase C).}
\label{tab:utias-case}
\end{table}

Two mixtures are considered in this section:  a  multi-temperature mixture composed of 3 species denoted by the set 
$\mathcal{S}_3 = \{\mathrm{Ar}$, $\mathrm{Ar}^+\!\!\!$, $\mathrm{e}^-\}$, so-called ``argon-3," and  a state-to-state mixture composed of 34 species denoted by the set $\mathcal{S}_{34} = \{ \mathrm{Ar(i)} ~|~(i=1,\ldots,31)$, $\mathrm{Ar}^+(1)$,  $\mathrm{Ar}^+(2)$, $\mathrm{e^-}\}$, so-called ``argon-34." The initial composition is assumed to be in local thermodynamic equilibrium at the freestream conditions.

The solution  is  recomputed in the post-shock region by means of the Lagrangian reactor using a state-to-state  model, starting from a two-temperature ($T$-$T^e$) baseline solution, where the translational energy of the heavy-particle bath is at temperature $T$, and the translational energy of electrons and electronic energy of the heavy-particle bath  at temperature $T^e$, the populations of the electronic energy levels of  $\mathrm{Ar}$ and $\mathrm{Ar}^+$  following a Boltzmann distribution. The post-shock conditions used for the baseline solution assume that the electron-electronic mode is frozen, so that temperature $T$ jumps, while temperature $T^e$ remains at its freestream value.  The baseline solution is obtained for the argon-3 mixture by solving in steady conditions the system \eqref{eqMaxT} coupled to eq.~\eqref{eqEM} for the mode at temperature $T^e$. The two temperatures then relax towards each other and eventually equilibrate at the same value. The state-to-state Lagrangian  eqs.~\eqref{eq:testcase-A-mass}, \eqref{eqAp3}, and \eqref{eqAp4} presented in Appendix are  solved using the argon-34 mixture. Note that two temperatures are considered in the state-to-state case: the heavy-particle translational temperature $T$ and the electron translational temperature $T_e$. A reference solution is obtained for the argon-34 mixture by solving in steady conditions the system \eqref{eqMaxT} coupled to the electron energy eq.~\eqref{eqEE}. 

The chemical mechanism and rate coefficients for the argon-34 mixture follow Kapper and Cambier's recommendations. \cite{kapper_ionizing_2011} 
The chemical mechanism for the argon-3 mixture comprises electron-impact ionization~\cite{annaloro_vibrational_2014}
and atom-impact ionization.
The forward rate coefficient $k_f(T^e)=A~(T^e)^n\exp(-T_0/T^e)$ for  argon-impact ionization is obtained by integrating experimental cross-section data\cite{hayden_low-energy_1966} and is subsequently multiplied by a scaling factor to retrieve the thermal equilibrium point of the reference state-to-state temperature profile, as reported in Table \ref{tab:forward-rate-cooked-s2s}.
Backward rate coefficients are computed from the equilibrium constant imposing micro-reversibility. 

\begin{figure}[htpb]
  \centering
  \includegraphics[width=1.0\columnwidth]{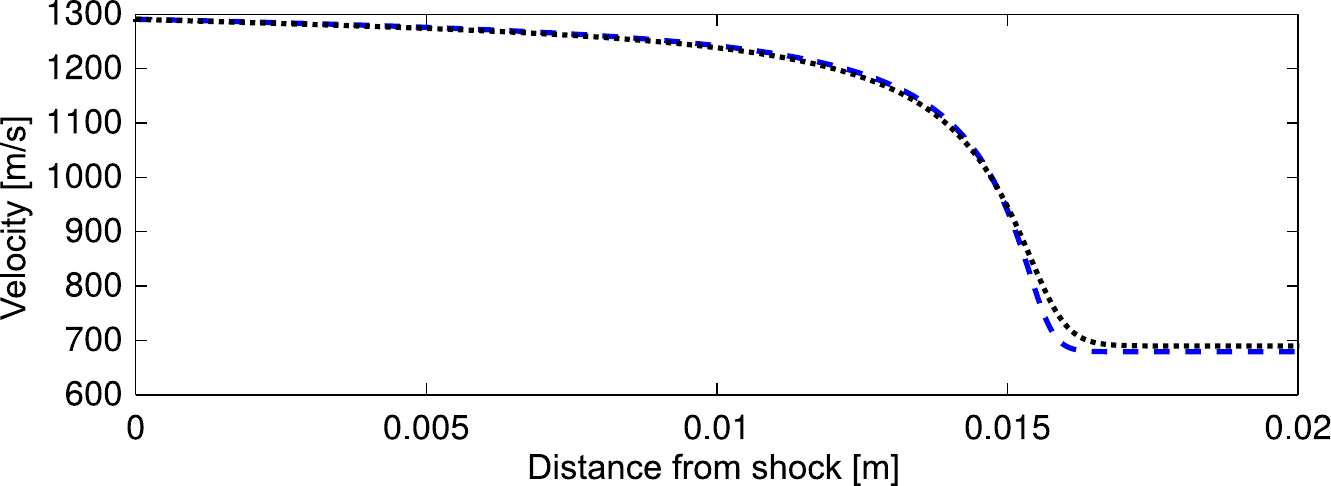}
  \caption{State-to-state refinement for UTIAS-Testcase C: 
  velocity in the post-shock region; 
   \textcolor{blue}{\bf -\,-\,-} baseline simulation (argon-3); $\bm{\cdots}$ reference simulation (argon-34).}
  \label{fig:s2s-modified-rate-velocity}
\end{figure}

\begin{table}
  \centering
  \begin{tabular}{ccc}
$A$ [$\mathrm{cm}^3/\mathrm{s}\,\mathrm{mol}$] & $n$ [--] & $T_0$ [K] \\
     \hline
$1.8247 \times 10^{-6}$ & $3.597$ & 69\,940 
  \end{tabular}
  \caption{Arrhenius parameters for forward rate coefficient of argon atom-impact ionization reaction.}
  \label{tab:forward-rate-cooked-s2s}
\end{table}

\begin{figure}[htpb]
    \centering
    \includegraphics[width=\columnwidth]{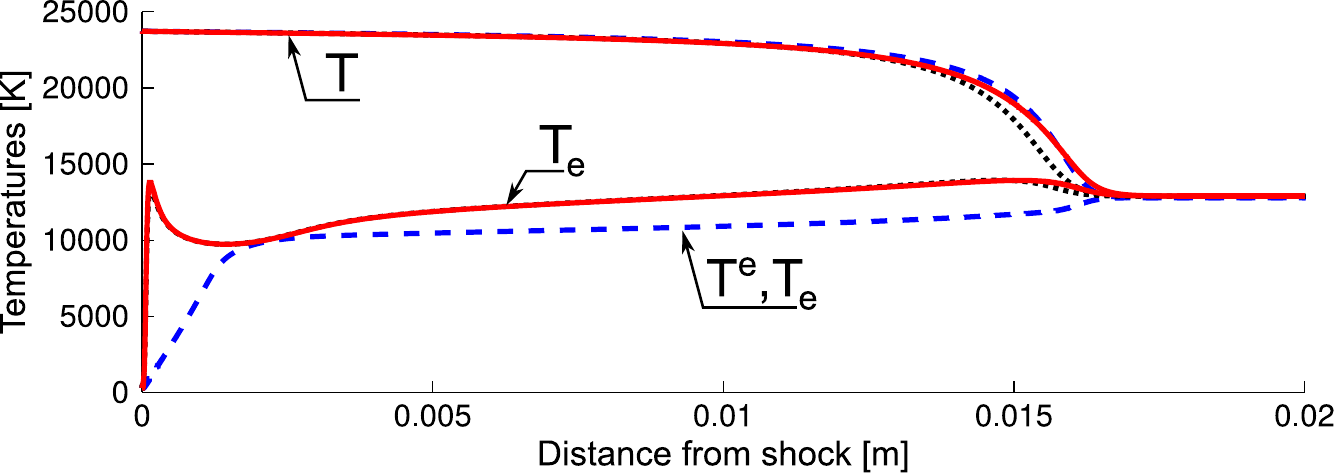}
    \caption{
    State-to-state refinement for UTIAS-Testcase C: heavy-particle translational temperature, electronic temperature and electron translational temperature in the post-shock region; 
     \textcolor{blue}{\bf -\,-\,-} baseline simulation (argon-3); \textcolor{red}{\bf -----} Lagrangian simulation (argon-3 $\rightarrow$ argon-34); $\bm{\cdots}$ reference simulation (argon-34).}
    \label{fig:s2s-T-Te}
  \end{figure}

The reference state-to-state flowfield is matched with good accuracy, as shown in Fig.~\ref{fig:s2s-modified-rate-velocity}.
The baseline solution is recomputed using the Lagrangian reactor with the state-to-state approach. 
Owing to the fidelity of the baseline flowfield, the Lagrangian simulation provides temperatures very close to the reference solution (Fig.~\ref{fig:s2s-T-Te}) and a complete matching in terms of argon electronic energy population.
In fact, if the \textit{exact} flow velocity and density would be used as an input for the Lagrangian solver, the exact solution would be retrieved, by consistency of the equations.

\subsection{2D rarefied flows}
We study  the excitation and ionization 
of an argon plasma past a bow shockwave in the rarefied regime in two parts. In Part 1, as a proof of concept, the Lagrangian solver is first used  on a baseline simulation  obtained by means of the Direct Simulation Monte Carlo  method for a single-species argon gas without considering its internal energy.
The first goal is to test the Lagrangian approach against a multidimensional flow around a cylinder, using the decoupled single-streamline mode introduced in section \ref{subsec:sol-Lagrangian-eq}.
The second goal is to demonstrate that the solver (which is hydrodynamic in nature)
can tackle the rarefied regime, importing enthalpy variations from the baseline simulation. In Part 2, the Lagrangian solver is run  for the argon-34 mixture using a collisional-radiative mechanism.

\begin{figure}[htb]
  \centering
  \includegraphics[width=1.0\columnwidth]{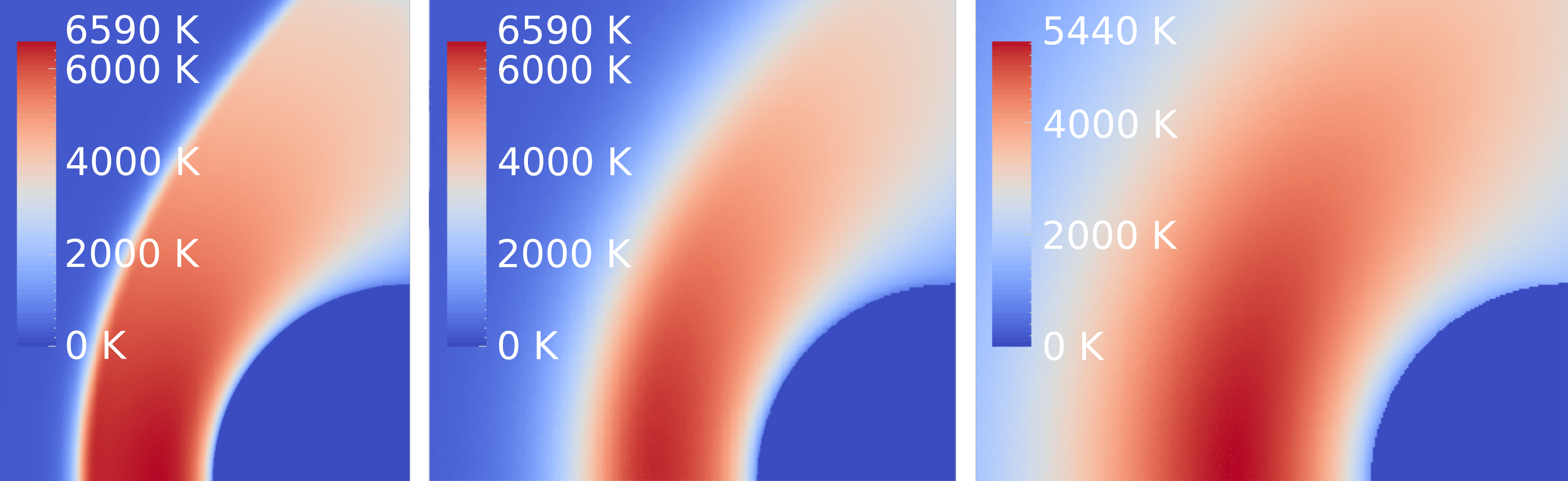}
  \caption{Argon flow around cylinder (Testcase D), baseline DSMC temperature field. Left: case (a) Kn =0.01; middle: case (b) Kn=0.05; right: case (c) Kn=0.25.}
  \label{fig:dsmc-Kn}
\end{figure}

The DSMC method 
\cite{bird_molecular_nodate} is a particle-based  approach for  solving the 
Boltzmann equation.
The result is thus valid even at high rarefaction degrees.
Additionally, the stochastic nature of DSMC provides an independent verification of our Lagrangian tool based on a (deterministic) partial differential equation formulation.
Argon is an inert gas chosen to test the ``rarefaction capabilities'' of the Lagrangian solver,
with no interference of the chemical refinement procedure. Three baseline solutions are obtained at different Knudsen numbers in the transition regime, using the DSMC software SPARTA. \cite{gallis_direct_2014}
Table~\ref{tab:fs-rarefied-cylinder} shows the freestream velocity $V_\infty$,  
temperature $T_\infty$, Mach number $\rm{Ma}_\infty$, body diameter $d$,  and wall temperature $T_{wall}$ taken from Lofthouse.\cite{lofthouse_nonequilibrium_2008} 
The Knudsen numbers based on a cylinder of diameter $d$ are obtained by adopting different freestream number densities, as shown in Table~\ref{tab:fs-rarefied-cylinder-kn-n}.
Baseline simulations temperature fields are shown in Fig.~\ref{fig:dsmc-Kn}. For a constant Mach number, the shock progressively exhibits a more diffuse behavior for higher values of the Knudsen number.

\begin{table}
\centering
\begin{tabular}{cccccc}
\hphantom{a} & $V_\infty$ [m/s] & $T_\infty$ [K]  & $\rm{Ma}_\infty$ [--] & $d$ [m] & $T_{wall}$ [K] \\
\hline
$i)$ & 2\,624 & 200  & 10 & 0.304 & 500 \\
$ii)$ & 6\,585 & 300  & 25 & 0.304 & 1500 \\
\end{tabular}
\caption{Freestream conditions for argon flow around cylinder (Testcase D). $i)$: conditions for Part 1. $ii)$: conditions for Part 2.}
\label{tab:fs-rarefied-cylinder}
\end{table}

\begin{table}
\centering
\begin{tabular}{ccc}
 & $\mathrm{Kn}_d$ [--]      &    $n_\infty$ [1/m$^{3}$] \\
\hline
case (a)  & 0.01     &    $4.247 \times 10^{20}$ \\
case (b)  & 0.05     &    $8.494 \times 10^{19}$ \\
case (c)  & 0.25     &    $1.699 \times 10^{19}$ \\
\end{tabular}
\caption{Knudsen numbers and number densities for argon flow around cylinder (Testcase D).}
\label{tab:fs-rarefied-cylinder-kn-n}
\end{table}

\begin{figure}
  \centering
  \includegraphics[width=1.0\columnwidth]{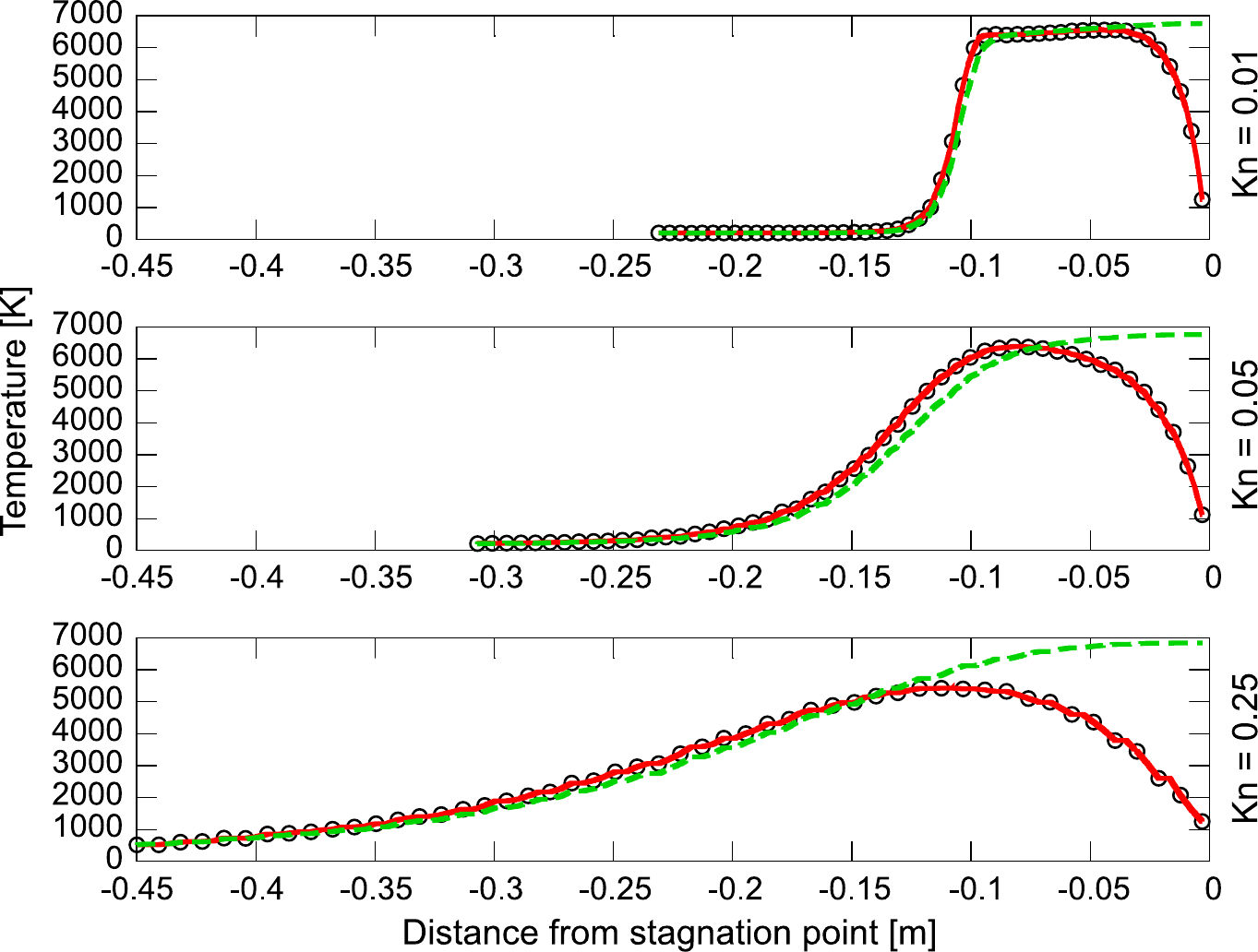}
  \caption{Argon flow around cylinder (Testcase D), temperature along stagnation line; $\circ\!\circ\!\circ$ baseline DSMC simulation; \textcolor{mygreen}{\bf -\,-\,-} adiabatic Lagrangian simulation; \textcolor{red}{\bf -----} Lagrangian simulation using baseline enthalpy dissipation.}
  \label{fig:dsmc-temp-stagline-yesnoQ}
\end{figure}

{\bf Part 1.} Starting from the baseline solution, the velocity and density fields are extracted
along the stagnation line and fed to the Lagrangian reactor. The Lagrangian eqs.~\ \eqref{eq:testcase-A-mass} and \eqref{eq:testcase-D-T-part1} given in Appendix are solved to recompute the temperature as a verification step.
Two alternative cases are studied: an adiabatic Lagrangian simulation obtained by setting the term $\Delta H^*/\Delta s = 0$ in eq.\ \eqref{eq:testcase-D-T-part1}, and 
a Lagrangian simulation using the baseline enthalpy dissipation
obtained by importing this term from the baseline solution.
In the former case, 
the Lagrangian solver cannot reproduce the temperature field close to the wall, as shown in Fig.~\ref{fig:dsmc-temp-stagline-yesnoQ}.
Due to the purely hyperbolic nature of the adiabatic formulation, the marching approach updates the solution based only on previous streamline points, and contains no information over the upcoming wall temperature.
Instead, the adiabatic simulation allows us to retrieve the total temperature value at the stagnation point.
This result complies with the nature of adiabatic flows, where there can be no energy exchange with surfaces.
When the heat flux is imported from the baseline simulation, the correct solution can be retrieved exactly.
Analogous simulations were performed on side streamlines other than the stagnation line, retrieving the very same degree of accuracy.

This testcase shows that the solver behaves well in rarefied multidimensional single-species flows.
The capability of importing the enthalpy from the baseline solution allows us to correctly reproduce heat fluxes even in rarefied conditions, as long as they are considered consistently with the simulated regime.
Likewise, this capability proves to be crucial to study continuum flows with temperature gradients, where it is important to reproduce the heat exchange with nearby fluid 
elements.

As long as specific closures are not introduced for the diffusion terms,
the governing equations solved with the Lagrangian approach are  general and fully equivalent
to mass, momentum and energy balances.
Their range of validity over different rarefaction regimes thus depends on the validity of 
energy fluxes themselves.
By importing the energy flux from the baseline DSMC solution, we can guarantee that they comply
with the rarefaction degree of the case considered. In this testcase, since no new species are introduced in the computation, the DSMC fluxes are
not only accurate, but numerically exact, giving a perfect matching of the solutions.
A different approach consists in \textit{computing} the heat fluxes, using for example Fourier's law,
valid in the continuum approach, as it will be detailed in Testcase E. In a multi-species flow, not only energy but also mass diffusion needs to be modeled:
importing the diffusive mass fluxes from the baseline solution is not feasible when chemical refinement is performed, since it introduces additional chemical species whose diffusion velocity is  unknown a 
priori.
An option is using continuum modeling for the mass fluxes, as shown in Testcase~E.

\begin{figure}[htpb]
  \centering
  \includegraphics[width=1.0\columnwidth]{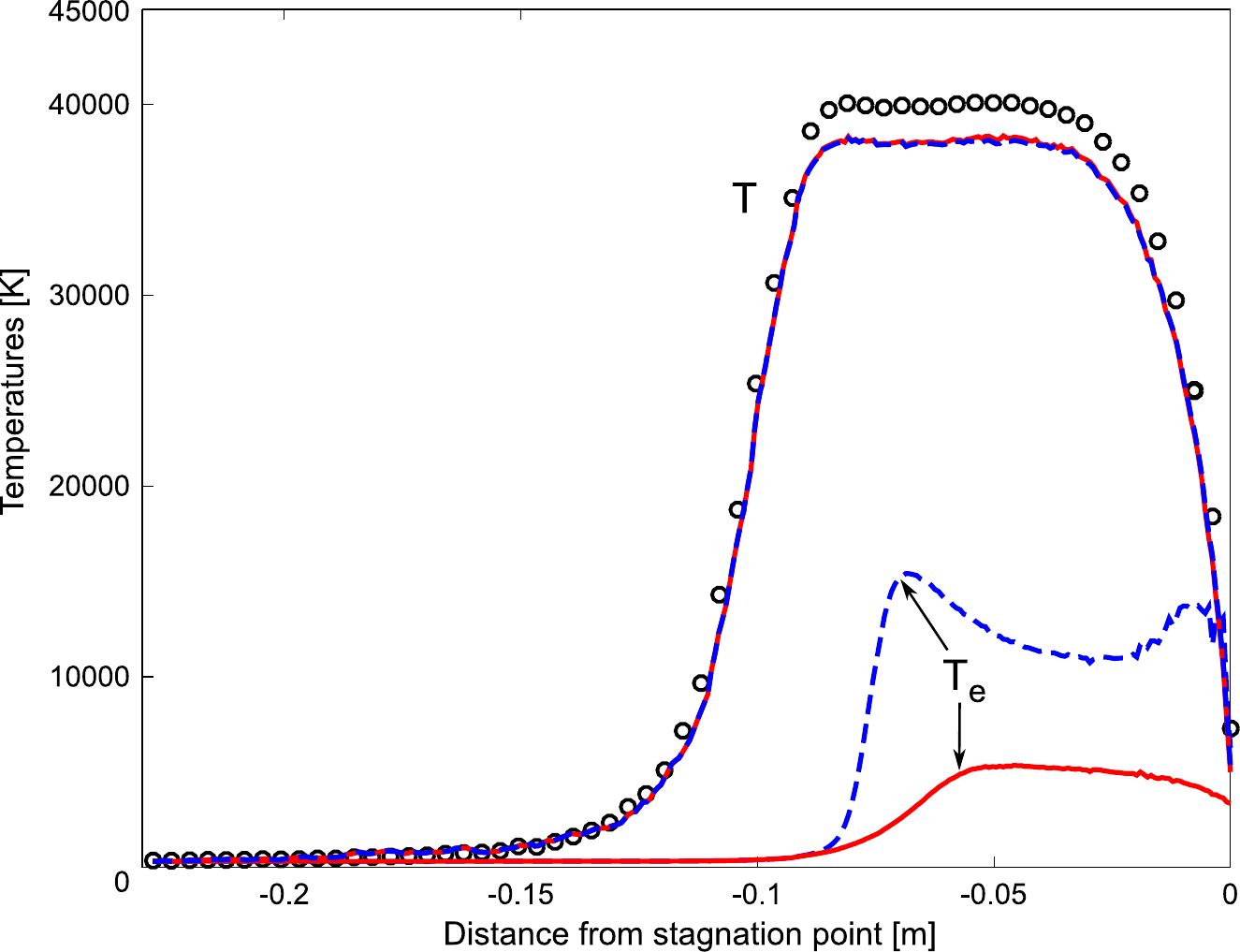}
  \caption{Argon flow around cylinder (Testcase D, Kn $=0.01$), heavy-particle and  electron translational temperatures along the stagnation line. 
  $\circ\!\circ\!\circ$ Baseline DSMC solution;
  \textcolor{blue}{\bf -\,-\,-} radiation disabled;
           \textcolor{red}{\bf -----} radiation enabled.}
  \label{fig:testcase-D-T-Te-rad-norad}
\end{figure}

\begin{figure}[htpb]
  \centering
  \includegraphics[width=1.0\columnwidth]{
  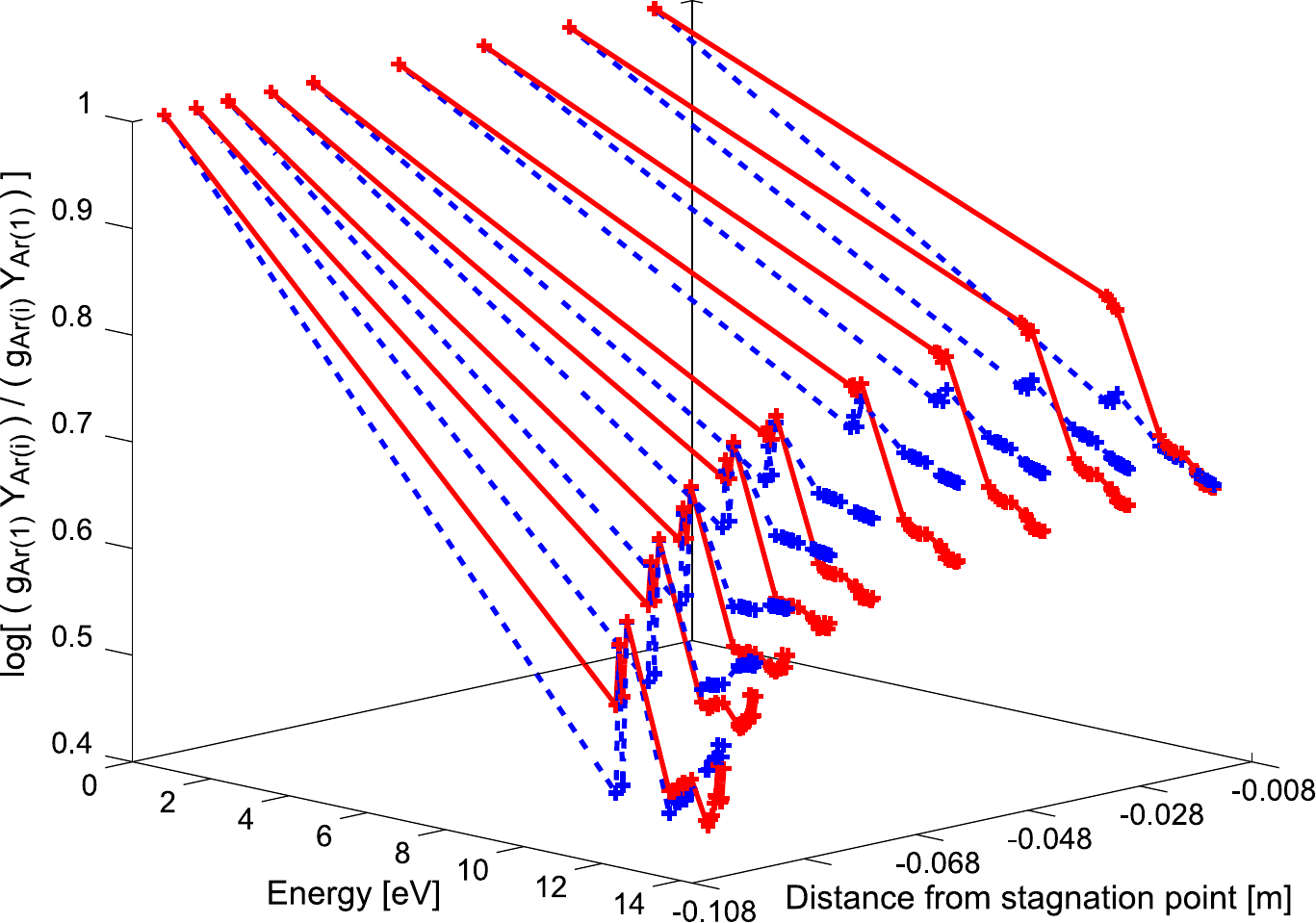}
  \caption{Argon flow around cylinder (Testcase D, Kn $=0.01$),  state-to-state population of argon electronic energy levels (logarithmic scale) normalized with respect to ground state
           along the stagnation line. 
           \textcolor{blue}{\bf -\,-\,-} Radiation disabled;
           \textcolor{red}{\bf -----} Radiation enabled.
           }
  \label{fig:testcase-D-relative-populations}
\end{figure}

{\bf Part 2.} At this point, a further computation is performed, to mimic a possible application.
A DSMC computation is performed at $\rm{Kn} = 0.01$ at higher freestream velocity, see Table~\ref{tab:fs-rarefied-cylinder}, in order to reach higher post-shock temperatures.
The computation is refined with a state-to-state approach, additionally enabling radiation compared to Testcase C.
The approach  consists in solving mass eq.\ \eqref{eq:testcase-A-mass}, electron temperature eq.\ \eqref{eq:testcase-D-Te-part2} and temperature eq.\ \eqref{eq:testcase-D-T-part2} like in Testcase C.
Total enthalpy variation is the only diffusive process accounted for, by importing quantity $\Delta H^* / \Delta s$ from the baseline DSMC computation.
The details of the source and radiation terms can be found in Kapper and Cambier.\cite{kapper_ionizing_2011}
The temperature profiles are shown in Fig.\ \ref{fig:testcase-D-T-Te-rad-norad}.
The DSMC simulation accounts for ground state argon only: 
this explains the difference in translational temperature between the baseline and Lagrangian simulations, where part of the energy is transfered to the excited electronic energy levels and the free electrons.
In Fig.~\ref{fig:testcase-D-relative-populations}, the populations of the argon electronic energy levels are shown along the stagnation line. They are normalized with respect to ground state as follows
\smash{$\log[( g_{{\rm Ar}(1)}Y_{{\rm Ar}(i)})/(g_{{\rm Ar}(i)}Y_{{\rm Ar}(1)} )]$}, where quantity $g_{{\rm Ar}(i)}$ stands for the degeneracy of level $i$.
Populations are seen to be non-Boltzmann. The effect of radiation is apparent for this case: the non\textcolor{magenta}{-}radiative case reaches equilibrium quicker along the shock, while radiation lowers the proportion of highly excited argon atoms, in favor of lower ones. It is clear that by enabling radiation, higher energy levels decay into lower energy 
ones, in the high-temperature post shock region.

\subsection{Mass and energy diffusion}
We study  ionization-recombination processes in air for the rarefied regime, as well as mass and energy diffusion of electrons,  along the trail past a blunt body moving at hypersonic speed. This  is the most inclusive among the testcases studied, combining chemical refinement with rarefaction and diffusion capabilities in a multidimensional flow.
Considering a 2D axisymmetric flow past a spheric body, mass and energy fluxes are \textit{modeled} 
in the transversal direction, allowing us to assess the coupled multi-streamlines mode introduced in section \ref{subsec:sol-Lagrangian-eq}. 
Table~\ref{tab:fs-hypersonic-sphere} shows the freestream velocity $V_\infty$, pressure $P_\infty$,
temperature $T_\infty$, Mach number $\rm{Ma}_\infty$,  body diameter $d$, and wall temperature $T_{\mathrm{wall}}$. 
The main idea behind this testcase is to include in the Lagrangian simulation elementary processes absent from the DSMC simulation, $i.e.$, reactions of   electron-ion recombination. 

\begin{table}
\centering
\begin{tabular}{cccccc}
$V_\infty$ [m/s] & $P_\infty$ [Pa] & $T_\infty$ [K] & $\rm{Ma}_\infty$ [--] & $d$ [m] & $T_{\mathrm{wall}}$ [K] \\
\hline
20\,000 &  3.61 & 220 & 67.3 & 0.01 & 2000 \\
\end{tabular}
\caption{Freestream conditions for partially  ionized air trail past a blunt body (Testcase E).}
\label{tab:fs-hypersonic-sphere}
\end{table}

\begin{figure}
  \centering
  \includegraphics[width=0.9\columnwidth]{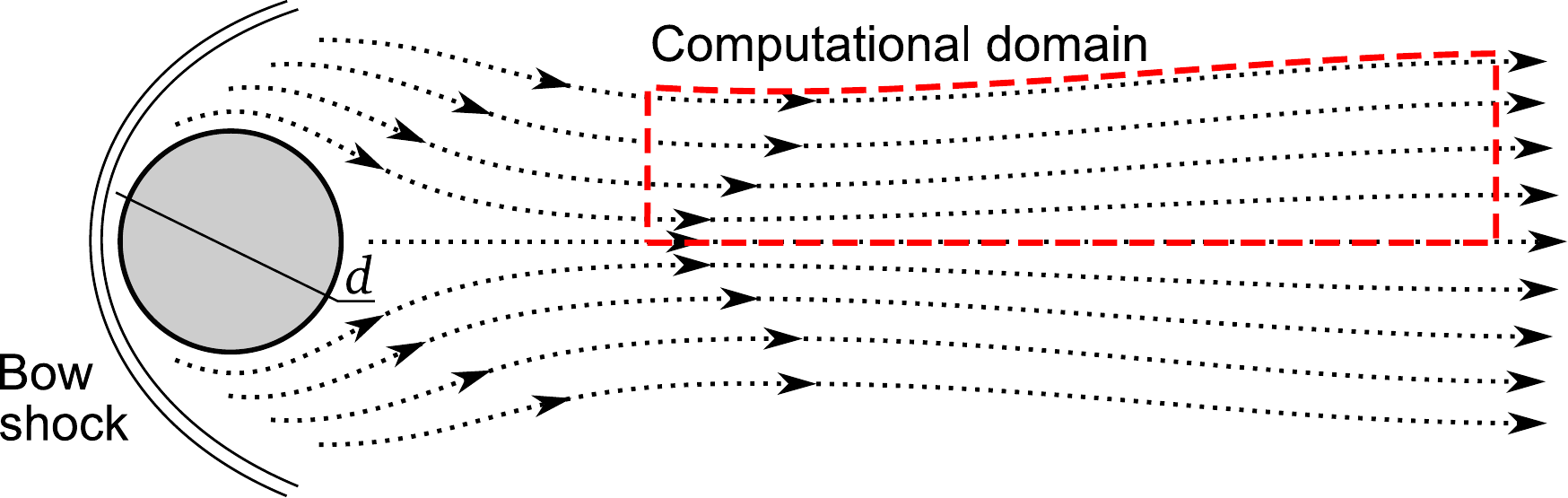}
  \caption{Computational domain for blunt body trail problem (Testcase E); out of scale.}
  \label{fig:meteor-domain}
\end{figure}

\begin{figure}
  \centering
  \includegraphics[width=1.0\columnwidth]{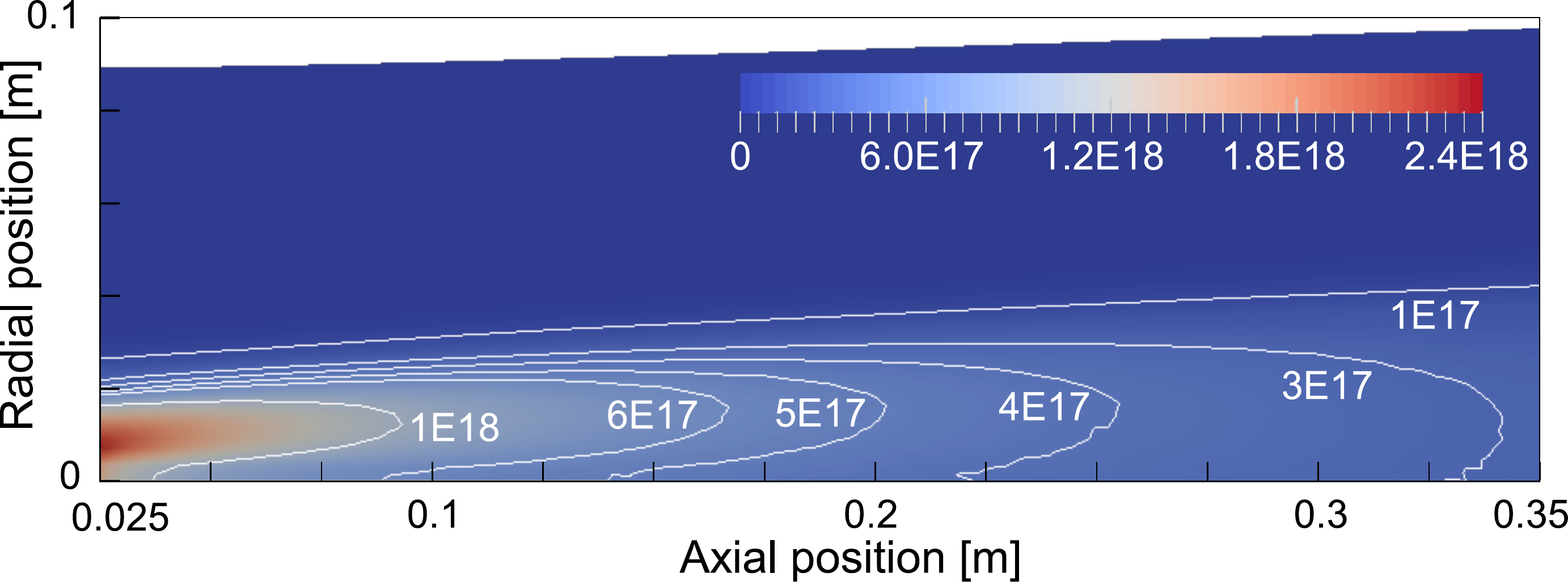}
  \caption{Partially ionized air trail past a blunt body (Testcase E): electron number density field for the Lagrangian simulation. Recombination included.}
  \label{fig:free-electrons-nrho-trail}
\end{figure}

Considering a freestream Knudsen number $\mathrm{Kn}\approx 0.1$ (based on the body diameter), the baseline solution is obtained  using the SPARTA software, for a chemically reacting air-11 mixture.
The DSMC method used inherently includes mass and energy diffusion but the chemical mechanism employed does \textit{not} take into account recombination reactions. First, we present  a verification step by recomputing the composition by means of the Lagrangian solver using the same chemical mechanism as in the baseline simulation, excluding recombination.
A number of streamlines are extracted from the baseline solution in the trail region. 
The computational domain, sketched in Fig.~\ref{fig:meteor-domain}, starts $2$ diameters after the rear stagnation point and extends for $32.5$ diameters. The upper side is limited by the upper streamline and the lower by the axis (coinciding with the lower streamline). Dimensions are reported in Fig.\ \ref{fig:free-electrons-nrho-trail}. 
 The eqs.\ \eqref{eq:testcase-E-mass} and \eqref{eq:testcase-E-temperature} solved are given in Appendix.
Transverse diffusion is modeled by means of a continuum closure: multicomponent diffusion for the species mass fluxes and  Fourier's law together with enthalpy diffusion for the  mixture heat flux.\cite{giovangigli_multicomponent_1999}
The simulation is based on the coupled multi-streamlines mode, picking the initial conditions from the first point of each streamline.
A number of $25$ streamlines was found to be enough to reproduce accurately the baseline electron concentration. 

\begin{figure}
  \centering
  \includegraphics[width=0.77\columnwidth]{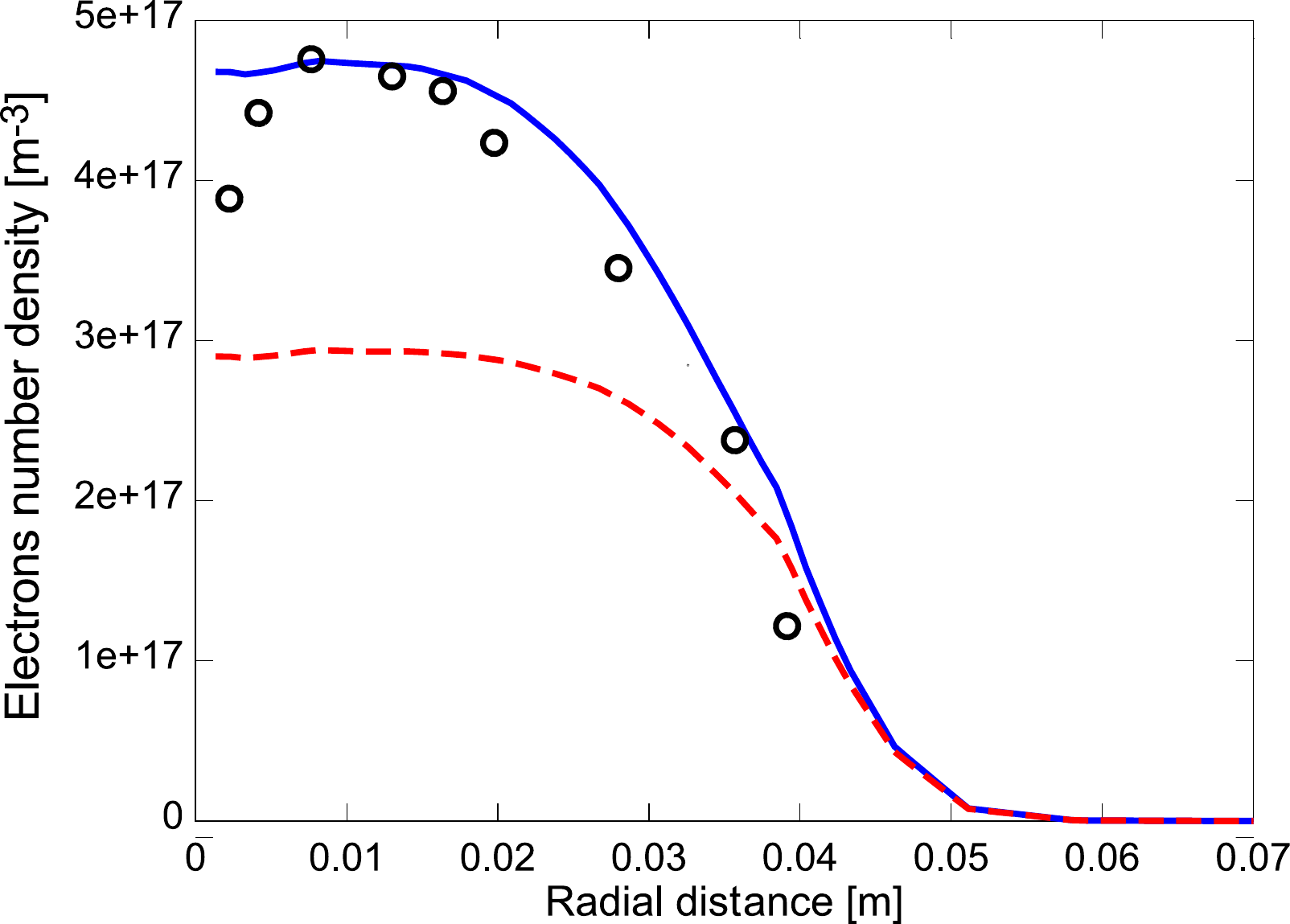}
  \caption{Partially ionized air trail past a blunt body (Testcase E): radial electrons number density at 0.35~m; $\circ\circ\circ$ baseline DSMC simulation;  \textcolor{blue}{\bf -----} Lagrangian simulation without recombination; \textcolor{red}{\bf -\,-\,-} Lagrangian simulation with recombination.}
  \label{fig:free-electrons-nrho-exit}
\end{figure}

The radial electron density  at the exit of the domain is compared to the baseline simulation in Fig.\ \ref{fig:free-electrons-nrho-exit}, showing a good agreement. The statistical scatter present in the DSMC solution is mainly due to the low density of free electrons, considering that the ionization degree at the exit is of the order of $0.01\%$.
In principle, a longer sampling time would reduce the scatter.
Note also that the noise is higher near the axis, due to the strong reduction of the cell size in axisymmetric configuration, resulting in even less simulated particles. The near-axis outlier points are to be considered a numerical artifact.

\begin{figure}
  \centering
  \includegraphics[width=1.0\columnwidth]{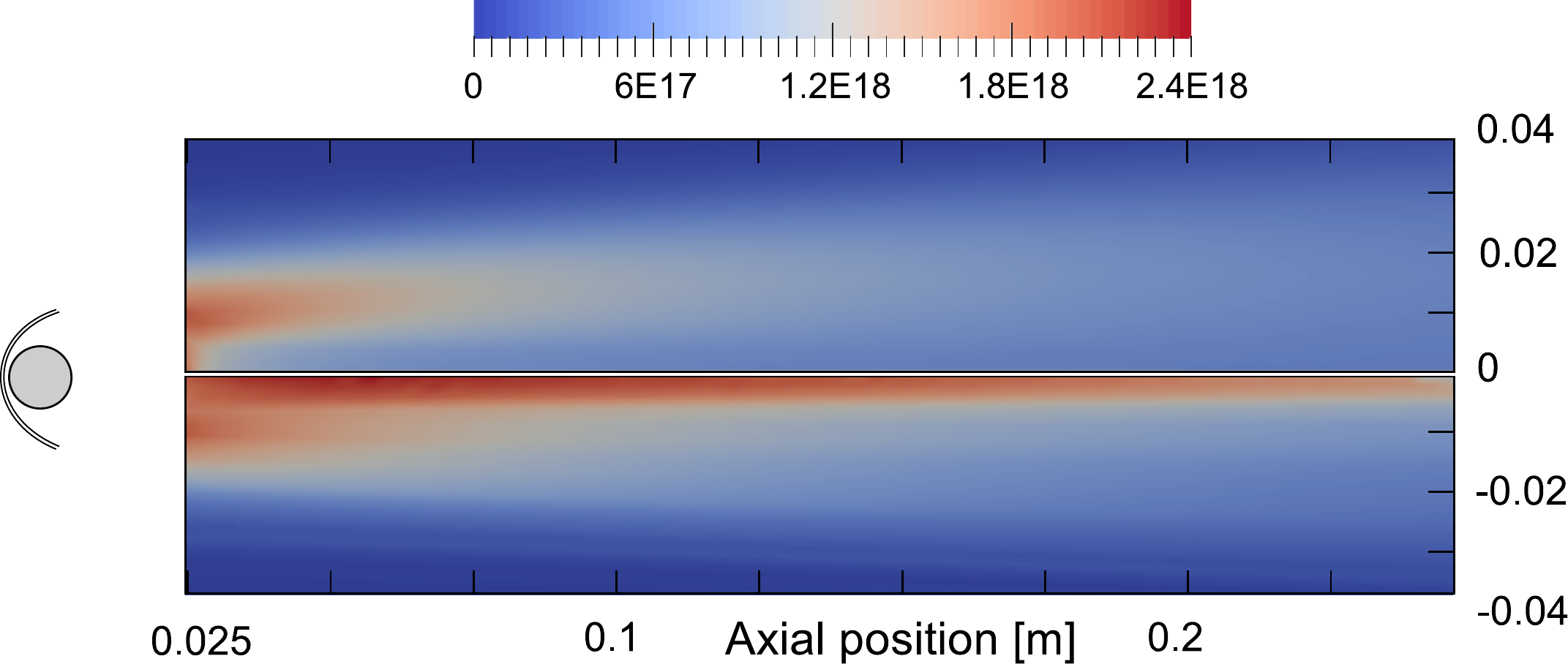}
  \caption{Partially  ionized air trail past a blunt body (Testcase E): effect of diffusion of free electrons versus recombination reactions. Electron number density [m$^{-3}$] for the Lagrangian simulation: top, only diffusion; bottom, only recombination.  Meteoroid included in scale for dimensional comparison.}
  \label{fig:free-electrons-diff-only-rec-only}
\end{figure}

The Lagrangian solver is then applied to include  recombination reactions a posteriori  in the chemical mechanism. The recomputed number density field of free electrons is reported in Fig.\ \ref{fig:free-electrons-nrho-trail}. As expected, the electron number density is  significantly reduced, as shown in Fig.~\ref{fig:free-electrons-nrho-exit}.
The effect of free electrons diffusion over recombination can be easily assessed, by enabling or disabling the respective terms in the equations.
Figure~\ref{fig:free-electrons-diff-only-rec-only} shows that diffusion prevails over recombination at these conditions as quantified by computing the third Damk\"ohler number, ratio of the diffusive-over-reactive timescale. Its value is  $\rm{Da_{\rm{III}}}= \tau^{\mathrm{diff}}/\tau^{\mathrm{react}}\approx 10^{-8}$.

Despite the degree of rarefaction, particles have enough time to thermalize moving along the trail. For this reason, recomputing fluxes using the continuum assumption works well in this testcase. In fact, sampling the velocity distributions function from the DSMC simulation gives quasi-Maxwellian shapes.
The Knudsen number in this problem would indeed result much smaller if, as a reference length, one would choose the distance along the axis.
The Lagrangian approach brings a number of advantages. 
First, the numerical scatter of minor species is reduced, naturally improving the accuracy of the chemical reactions prediction (and radiation if included). 
Second, one could perform lighter DSMC simulations by reducing the number of species and performing a computationally efficient refinement a posteriori with a large number of species. Particular reactions not yet implemented in a DSMC code can be easily introduced as well.

\begin{figure}
  \centering
  \includegraphics[width=1.0\columnwidth]{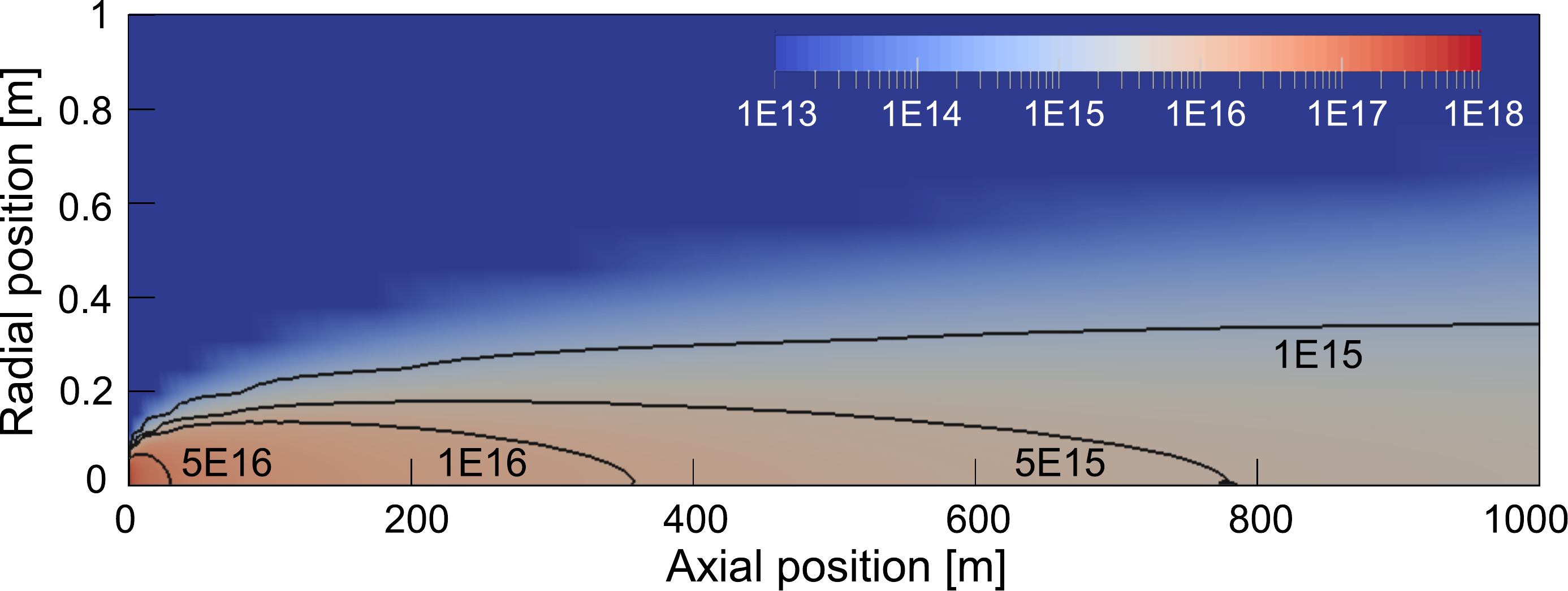}
  \caption{Partially  ionized air trail past a blunt body (Testcase E): electron number density [m$^{-3}$] for the Lagrangian simulation based on extrapolated baseline simulation in a 1~km trail. Axis not in scale.}  \label{fig:free-electrons-longtrail}
\end{figure}

Finally, regarding blunt body trail simulations, it is possible to use the Lagrangian reactor to obtain solutions up to extremely long distances from the body. In fact, in rarefied conditions velocity and density often reach freestream values, while free electrons are still not totally recombined. This feature
allows us to obtain baseline solutions that would be too expensive to achieve by means of the DSMC method:
the trail is extended assuming straight streamlines with freestream values for velocity and density.
Taking the results obtained above as an initial condition, the Lagrangian solver was used to study the evolution of free electrons in the trail up to $0.05$ seconds after the body passage, equivalent to a trail length of 1~km, $i.e.\ 10\,000$ times the body diameter. Figure\ \ref{fig:free-electrons-longtrail} shows the persistence of  free electrons in the trail. 
The simulation by means of the Lagrangian solver takes around 20 minutes on a laptop.
On the other hand, a full DSMC simulation of the same domain would be \textit{extremely} heavy, requiring to simulate around $10^9$ particles and $250$ million cells in order to meet the DSMC convergence criteria.\cite{bird_molecular_nodate}

\section{Conclusions}\label{sec:conclusions}

In this work, we discussed the development and testing of a diffusive Lagrangian solver allowing us to compute the  thermochemical state of  reactive and plasma flows.
Starting from a baseline solution obtained with any analytical, numerical, or even experimental methods, the Lagrangian solver can be used to introduce an arbitrarily large number of chemical species and their related chemical mechanism, together with a specific treatment of the internal energy based on multi-temperature or state-to-state models. The governing equations are solved along the streamlines, importing the velocity and density fields  directly from the baseline solution, carrying the information over the original problem geometry. The Lagrangian solver recomputes the chemical species concentration and the temperature fields, 
for both the single and multiple streamlines modes developed.
The single-streamline mode was formulated in such a way that dissipative fluxes of energy can be externally introduced in the formulation. This allowed us to extend the validity of the method to rarefied conditions by considering a baseline solution obtained by means of the DSMC method. For the multi-streamline mode, a  finite-volume method was proposed to compute the transverse mass and energy diffusive fluxes.

The  Lagrangian solver  was assessed for a number of testcases.
First, the chemical and thermal refinement capabilities were examined using multi-temperature models, studying the thermochemical relaxation past a 1D strong shockwave in an air mixture.
The test was repeated for an argon mixture, employing a state-to-state collisional-radiative model.
The approach was then tested on multidimensional cases, studying a 2D argon flow over a cylinder in the transition regime, then moving to the simulation of a chemically reacting air trail past a blunt body in 2D axisymmetric geometry, computing mass and energy diffusion across streamlines.

For all the testcases studied, the solver developed was used to estimate the effects of a detailed chemistry or thermal nonequilibrium on a baseline simulation. The results demonstrate that the method developed enables coupling detailed mechanisms to multidimensional solvers to study thermo-chemical nonequilibrium flows. 
One should notice that the imported velocity and density fields appear in the Lagrangian equations through the  mass flux $\rho U$ and  kinetic energy flux $U^3$.
While the former is constant along each streamline, the latter may change according to the thermochemical model adopted.
In particular, when minor species are introduced using the Lagrangian reactor, these species can be thought as transported scalars, not altering significantly the energy balance of the mixture.
The simplified velocity field is found to be very close to the one obtained for the full mixture, and the Lagrangian solver provides very accurate solutions.
More care should be paid when newly introduced minor species play an important role in the chemical mechanism, or when the coarse thermochemical model is too poor to describe correctly the baseline solution, in terms of hydrodynamic features.
The Lagrangian reactor can always be tested on simplified representative problems, to assess which accuracy can be expected for the real problem. Nonetheless, in all the testcases performed, the Lagrangian reactor improved drastically the baseline simulations. The computational cost of a Lagrangian recomputation is typically  orders of magnitude smaller with respect to a full solution of the problem.
The solver has the additional benefit of being immune from statistical noise, which strongly affects the accuracy of DSMC simulations, especially considering minor species in the mixture.

A number of applications can be tackled with the  method developed, ranging from the interpretation of measurement techniques such as optical emission spectroscopy to design of plasma generators.
The testcase performed suggests applications in the refinement of the temperature profile past shock waves, and the estimation of radiative properties employing collision-radiative models. 
The calculation of the  free electron concentration in hypersonic trails was also discussed in this paper, which can easily be applied to the fields of telecommunication blackout and the radio-detection of meteors  during atmospheric entry.\cite{wislez_forward_1996,lamy_home_nodate} Extended trail simulations of meteors is an ongoing work performed by means of the Lagrangian solver, including ablation products. \cite{bariselli_aerothermodynamic_2018}

\section*{Appendix}
We consider a system of elementary reactions $r\in\mathcal{R}$ formally written as
\begin{equation*}
 \sum_{i \in \mathcal{S}} \nu^{'}_{ir} X_i \rightleftharpoons \sum_{i \in \mathcal{S}} \nu^{''}_{ir} X_i,
\end{equation*}
where $X_i$ is the chemical symbol for species $i\in\mathcal{S}$,  and $\nu^{'}_{ir}$ and $\nu^{''}_{ir}$, the forward and backward stoichiometric coefficients for species $i$ in reaction $r$. The
species production rates, compatible with the law of mass action, are expressed as \smash{$
 \omega_i = M_i\sum_{r \in \mathcal{R}} \nu_{ir} \tau_{r}$},
where $\nu_{ir} = \nu^{''}_{ir}-\nu^{'}_{ir}$ and symbol $M_i$ stands for the species molar mass. The rate of progress for reaction $r$ is given by:
\begin{equation*}
\tau_{r} = k_{r}^f\prod_{i \in \mathcal{S}}\left(\frac{\rho_i}{M_i}\right)^{\nu^{''}_{ir}} - 
		   k_{r}^b\prod_{i \in \mathcal{S}}\left(\frac{\rho_i}{M_i}\right)^{\nu^{'}_{ir}}.
\end{equation*}

The {\bf mass conservation} equations  solved for the test cases of Section \ref{sec:results} are given as follows. \\

\noindent \textbf{Testcases A, B, C, and D}
\begin{equation}\label{eq:testcase-A-mass}
  \frac{\mathrm{d} Y_i}{\mathrm{d} s}
  =
  \frac{\omega_i}{\rho U}, \quad i \in \mathcal{S},
\end{equation}

\noindent \textbf{Testcase E}
\begin{equation}\label{eq:testcase-E-mass}
  \frac{\partial Y_i}{\partial x}+  
  \frac{1}{\rho U \, \cos{\alpha_k}}
  \frac{\partial J_i}{\partial r}
  =
  \frac{\omega_i}{\rho U\, \cos{\alpha_k}},
  \quad i \in \mathcal{S},
\end{equation}
where index $k$ refers to the streamline considered. The transverse mass flux is 
$J_i = \rho_i V_i$, where the diffusion velocity \smash{$V_i =   - \sum_{j \in \mathcal{S}}$} $ D_{ij} {\partial} X_j/{\partial r} + \rho_iq_i E/ (m_iP)$ depends on the transverse gradients of species mole fractions $X_i$, through  multicomponent diffusion coefficients $D_{ij}$, and the (ambipolar) electric field $E$, where $q_i$ is the  charge of species $i$ and $m_i$ its mass.\\

The {\bf energy conservation} equations  solved for the testcases of Section \ref{sec:results} are given as follows.\\

\noindent \textbf{Testcase A}\\
%

%
\begin{equation}\label{eq:testcase-A-temperature}
  \frac{\mathrm{d}T}{\mathrm{d}s} 
=
- \left[
  \tfrac{1}{3}\rho \frac{\mathrm{d} U^3}{\mathrm{d}s}
+ \sum_{j\in\mathcal{S}}
    {h_j\omega_j}
  \right] \Bigg/
  \left[{\rho U}
    \sum_{j \in \mathcal{S}} c_{p,j} Y_j
  \right],
\end{equation}

\vspace{1ex}
\noindent \textbf{Testcase B}\\
%
\begin{equation}
  \frac{\mathrm{d}T^{v}}{\mathrm{d}s} 
=
   \left[
   {\Omega^{v,e}}
  - \sum_{j\in\mathcal{H}}
    {e^{v,e}_j\omega_j}
  \right] \Bigg/
  \left[{\rho U}
    \sum_{j \in \mathcal{H}} c_{v,j}^{v,e} Y_j
  \right],\label{eqAp1}
\end{equation}
\begin{multline}
  \frac{\mathrm{d}T}{\mathrm{d}s} 
=
  - \left[
      \tfrac{1}{3}\rho \frac{\mathrm{d} U^3}{\mathrm{d}s}
    + \sum_{j\in\mathcal{H}}
      {h_j\omega_j}
  \right.\\
  \left.
  + 
      \left(
        \Omega^{v,e} 
  - \sum_{j \in \mathcal{H}} e_j^{v,e} \omega_j
    \right)\right]
  \Bigg/
  \left[{\rho U}
    \sum_{j \in \mathcal{H}} c_{p,j}^{t,r} Y_j
  \right], \label{eqAp2}
\end{multline}
where the  source term for the vibrational-translational energy  relaxation (Landau-Teller-Millikan-White) and chemistry-energy coupling is introduced as follows,\cite{parkbook} $\Omega^{v,e}=  \sum_{j \in \mathcal{H}}\rho_j[{e^v_j\left(T\right)-e^v_j\left(T^{v}\right)}]/{\tau_j^{VT} } 
    + 
    \sum_{j \in \mathcal{H}} e^{v,e}_j \omega_j
 $.

\vspace{1ex}

\noindent \textbf{Testcase C}\\
%
\begin{equation}
\label{eqAp3}
  \frac{\mathrm{d}T_{e}}{\mathrm{d}s} 
=
   \left[
   \Omega_{e}
 + \frac{P_e U}{\rho} \frac{\mathrm{d} \rho}{\mathrm{d} s} 
 - e^{t}_e\omega^e
  \right]  \Big/
  \left[\phantom{\hspace{-.5cm}\frac{\mathrm{d} \rho}{\mathrm{d} s}}{\rho U}
  c_{v,e} Y_e
  \right] ,
\end{equation}
\begin{multline}
\label{eqAp4}
  \frac{\mathrm{d}T}{\mathrm{d}s} 
=
  -\left[
     \tfrac{1}{3}\rho\frac{\mathrm{d} U^3}{\mathrm{d}s}
    + \sum_{j\in\mathcal{S}}
      {h_j\omega_j}
      + {\gamma_e}
        \left(
        \Omega_{e}
      \vphantom{\frac{P_e U}{\rho}}
      + \frac{P_e U}{\rho} \frac{\mathrm{d} \rho}{\mathrm{d}s}
       \right.
      \right.
\\
  \left.
    \left.
      \vphantom{\frac{\mathrm{d} \rho}{\mathrm{d}s}}
    - e_e^{t} \omega_e
    \right) \vphantom{\sum_j}
  \right]\Bigg/
  \left[ {\rho U}
    \sum_{j \in \mathcal{H}} c_{p,j}^{t} Y_j
  \right],
\end{multline}
where the  source term for the electron heavy-particle translational energy  transfer and chemistry-energy coupling is introduced as follows,\cite{parkbook} 
{$\Omega_{e}=\rho_e [{e^T_e\left(T\right)-e^T_e\left(T_{e}\right)}]/{\tau^{ET}} + 
 \sum_{r \in \mathcal{R}^{\star}} \Delta h^r \tau^{r}
 + e^{t}_e\omega^e
$}. The set $\mathcal{R}^{\star}\subset \mathcal{R}$ comprises the electron-impact ionization and excitation reactions.

\noindent \textbf{Testcase D}\\

\noindent {\bf Part 1:} Thermal equilibrium case

\begin{equation}\label{eq:testcase-D-T-part1}
  \frac{\mathrm{d}T}{\mathrm{d}s} 
=
  \left[\vphantom{  \sum_{j}}
  \frac{\Delta H^*}{\Delta s}
- \tfrac{1}{2}\frac{\mathrm{d} U^2}{\mathrm{d}s}
  \right] \Bigg/
  \left[
   \sum_{j \in \mathcal{S}} c_{p,j} Y_j 
  \right],
\end{equation}

\noindent {\bf Part 2:}  Collisional-radiative model case

\begin{equation}\label{eq:testcase-D-Te-part2}
  \frac{\mathrm{d}T_{e}}{\mathrm{d}s} 
=
   \left[
   \Omega_{e}
  +\mathscr{P}_e
  + \frac{P_e U}{\rho} \frac{\mathrm{d} \rho}{\mathrm{d} s} 
  - e^{t}_e\omega_e
  \right]\Big/ \left[\rho U \, c_{v,e}^{t} Y_e\vphantom{\frac{P_e U}{\rho}}\right],
\end{equation}
\begin{multline}\label{eq:testcase-D-T-part2}
  \frac{\mathrm{d}T}{\mathrm{d}s} 
=
  \left[
      {\rho U}\frac{\Delta H^*}{\Delta s}
    + {\mathscr{P}}
    - \tfrac{1}{3}\rho\frac{\mathrm{d} U^3}{\mathrm{d}s}
    - \sum_{j\in\mathcal{S}}
      {h_j\omega_j}
      - {\gamma_e}
        \left(
        \Omega_{e}
      \phantom{\frac{P_e U}{\rho}}
    \right.
      \right.
\\
  \left.
    \left.
+ \mathscr{P}_e
      - \frac{P_e U}{\rho} \frac{\mathrm{d} \rho}{\mathrm{d}s}
    - e_e^{t} \omega_e
    \right)      \phantom{\sum_j}\hspace{-.5cm}
  \right]\Bigg/
  \left[\rho U
    \sum_{j \in \mathcal{H}} c_{p,j}^{t} Y_j
  \right],
\end{multline}
where $\mathscr{P}= - \sum_{(i,j)\in\mathcal{S}^{BB}} \Lambda_{ij} ( h_j^f - h_i^f) n_j A_{ji}$ expresses energy lost through bound-bound transitions from an upper electronic energy level $j$ of argon to a lower level $i$, denoted by the set $\mathcal{S}^{BB}$, with the escape factors assumed to be $\Lambda_{ij}=0$ for transitions to the ground state (optically thick) and $1$ for all the others (optically thin); and $A_{ji}$ the Einstein coefficient for the transition. 
The term \smash{$\mathscr{P}_e=\sum_{i\in\mathcal{S}_{31}} (h^f_{{\rm Ar}^+} - h^f_i) n_i R_i$} $ - 1.42\times10^{-40}Z_{\mathrm{eff}}^2 T_e^{1/2} n_{{\rm Ar}^+} n_e~[{\rm W}/{\rm m}^3]$ accounts for photorecombination with rate coefficient $R_i$, where $\mathcal{S}_{31} = \{ \mathrm{Ar(i)} ~|~(i=1,\ldots,31)\}$, and Bremsstrahlung with an effective charge $Z_{\mathrm{eff}}^2=1.67$.\cite{bellemans_simplified_2017}

\vspace{1ex}
\noindent \textbf{Testcase E}\\
\begin{multline}\label{eq:testcase-E-temperature}
  \frac{\partial T}{\partial x}
+
  \left[ \frac{\partial q}{\partial r}-
    \sum_{j \in \mathcal{S}} h_j
    \left(
      \frac{\partial J_j}{\partial r}
    \right)
  \right]  \Bigg/
  \left[
    \rho U \, \cos{\alpha_k} \sum_{j \in \mathcal{S}} c_{p,j} Y_j 
  \right]
\\
=-\left[
       \tfrac{1}{3} \rho \frac{\partial U^3}{\partial s}
    + \sum_{j \in \mathcal{S}} 
    h_j \omega_j
  \right]
  \Bigg/
  \left[
    \rho U \, \cos{\alpha_k} \sum_{j \in \mathcal{S}} c_{p,j} Y_j 
  \right],
\end{multline}
where $k$ is the index of streamline considered. The mass flux $J_i$, $i\in\mathcal{S}$ is defined as in eq.~\eqref{eq:testcase-E-mass}. The transverse heat flux is introduced as $q = - \lambda\frac{\partial}{\partial r}T
+ \sum_{j \in \mathcal{S}} h_j J_j$, where $\lambda$ is the thermal conductivity.


%
%

%

\section*{Acknowledgements}
The authors wish to acknowledge Dr.~Bellas-Chatzigeordis and Dr.~Bellemans (von Karman Institute for Fluid Dynamics) for the numerical support,
as well as Prof.~Frezzotti (Politecnico di Milano) for useful discussions on the topic, and Mr. Amorosetti (CORIA) for the careful reading of the paper.
The research of F.\ B.\ is funded by a PhD grant of the Research Foundation Flanders (FWO) and the one of B.\ D.\ by a PhD grant of the Funds for Research Training in Industry and Agriculture (FRIA). 
The research of T.\ M. \ was funded by the Belgian Research Action through Interdisciplinary Networks (BRAIN) funding CONTRAT BR/143/A2/METRO.



\end{document}